\documentclass[journal,comsoc]{IEEEtran}

\usepackage[numbers, sort&compress]{natbib}
\usepackage{graphicx}
\usepackage{epsfig}
\usepackage[dvipsnames]{xcolor}
\usepackage{tikz}
\usetikzlibrary{arrows.meta,positioning,calc,fit}

\tikzset{
	srdrBase/.style={
		draw=black!65,
		rounded corners,
		align=center,
		inner sep=3.2pt,
		minimum height=7.5mm,
		font=\scriptsize
	},
	srdrVar/.style={srdrBase, draw=MidnightBlue!70!black, fill=MidnightBlue!8},
	srdrLoad/.style={srdrBase, draw=BurntOrange!75!black, fill=BurntOrange!10},
	srdrDelay/.style={srdrBase, draw=ForestGreen!70!black, fill=ForestGreen!10},
	srdrRel/.style={srdrBase, draw=Purple!70!black, fill=Purple!9},
	srdrAux/.style={srdrBase, draw=black!55, fill=black!6},
	srdrArrow/.style={-{Latex[length=1.6mm]}, thick, draw=black!70},
	srdrDashed/.style={srdrArrow, dashed, draw=black!60}
}

\usepackage{orcidlink}
\hypersetup{hidelinks}

\usepackage{url}
\usepackage{enumerate}
\usepackage{booktabs}
\usepackage{multirow}
\usepackage{bm}
\usepackage{bbm}
\usepackage{commath}
\usepackage{amsfonts,amsthm,amssymb,amsmath}
\usepackage{algorithm}
\usepackage{algorithmic}
\usepackage{float}
\usepackage{comment}
\usepackage{color}

\graphicspath{{./Figures}}
\usepackage{subcaption}

\begin{document}

\title{RATIO: Redundancy-Controlled Stochastic Routing for Reliable Vehicular Multi-Hop Networking}

\author{Lei Lei$^{\orcidlink{0009-0009-0264-6044}}$,~\IEEEmembership{Student Member,~IEEE,}
	and Xudong Wang$^{\orcidlink{0000-0002-1353-1420}}$,~\IEEEmembership{Fellow,~IEEE}
	\thanks{
		Lei Lei is with Global College, Shanghai Jiao Tong University. 
		Xudong Wang is with The Hong Kong University of Science and Technology (Guangzhou).
		Corresponding author: Xudong Wang (e-mail: wxudong@ieee.org).
	}
}

%

\maketitle

	\begin{abstract}
	Reliable, low-latency multi-hop data delivery in vehicular networks is increasingly demanded, yet remains challenging due to frequent route failures caused by high mobility and intermittent blockage.
	While redundancy-based routing enhances robustness by forwarding packets over multiple paths, over-replication intensifies contention and introduces additional delay, highlighting the need to carefully managing redundancy--reliability trade-off.
	However, conventional deterministic multi-path replication typically duplicates packets to an integer number of branches, making the redundancy level hard to tune and adapt to time-varying network dynamics in vehicular networks.
	To this end, Redundancy-Controlled Stochastic (RATIO) routing is proposed in this paper.
	For each active flow, RATIO constructs a weighted reduced directed acyclic graph (DAG) as the routing structure, where edge weights specify per-link forwarding probabilities.
	At fork nodes, the aggregate outgoing forwarding probability is allowed to exceed one and a modulo-based stochastic forwarding rule is employed to guarantee feasible forwarding, thereby enabling continuously controllable redundancy.
	An idealized RATIO design is formulated as a load-minimizing optimization subject to per-flow timely-reliability and link-capacity constraints, but the problem is generally intractable under time-varying wireless dynamics.
	Accordingly, a practical heuristic, termed H-RATIO, is developed.
	H-RATIO constructs a compact reduced DAG by taking the union of candidate paths and optimizes forwarding probabilities via local scoring and replication-adjustment iterations.
	Extensive trace-driven SUMO/ns-3 co-simulations demonstrate that RATIO/H-RATIO consistently achieves the highest timely PDR compared to baselines, while providing substantially better delivery efficiency, especially under high-load scenarios.
\end{abstract}

\section{Introduction}
\label{sec:intro}

Recently, vehicular networks are increasingly expected to support reliable and low-latency delivery not only for safety-critical services,
but also for data-intensive traffic such as shared sensor streams, high-definition (HD) map updates, and multimedia content
\cite{2008_ComMag_VANET_tutorial,2014_JNCA_VANET_survey,2021_ACA_VANET_survey}.
Considering limited infrastructure coverage and intermittent connectivity caused by vehicle mobility, multi-hop networking is essential in vehicular networks
In practice, such services must often be delivered under intermittent, where data packets are relayed through intermediate vehicles towards their destinations
\cite{2020_AiminTang_Mesh, Mesh_V2I, Mesh_V2V}.

Due to high mobility and intermittent non-line-of-sight (NLOS) blockage, multi-hop transmissions in vehicular networks are prone to frequent route failures
In particular, a single broken link can trigger the route re-computation and recovery of the entire path, which introduces additional delay and packet loss, and can cause abrupt degradation of end-to-end delivery performance~\cite{2007_VTMag_routing_survey}.

To address this challenge, mobility-aware and stability-based routing policies have been proposed. 
These methods first predict link stability from vehicle motions and/or historical link-quality measurements, and then incorporate the resulting stability estimates into routing decisions to favor longer-lived paths~\cite{2010_PersComm_Stability_QoS,2015_NetCompApp_Stability_routing_survey}. In this way, route breakages can be reduced and the frequency of route recovery can be lowered. However, relying on link stability alone is often insufficient.
The end-to-end path still follows a single-route structure and thus remains inherently vulnerable to any hop breakage along the selected route.

Another widely used approach is to introduce redundancy.
A representative method is multi-path replication routing, where each packet is duplicated and sent along multiple paths towards the destination, thereby increasing the probability that at least one copy survives from transient path breakages~\cite{marina2001ondemand_multipath}.
Opportunistic forwarding, on the other hand, exploits the broadcast nature of wireless channels by allowing multiple candidate relays to participate the forwarding process and hence increases path diversity~\cite{biswas2005exor}.
However, redundancy is not free. Each additional copy occupies wireless channel resources, increases contention and collision probability, and may further enlarge queueing delay under finite network capacity.
Therefore, redundancy should not be used simply whenever multiple forwarding opportunities are available.
Instead, it should be carefully controlled according to network conditions and service requirements.
For example, under low mobility and high traffic load, a lower redundancy level is preferable to avoid unnecessary channel occupation and congestion.
In contrast, under high mobility and low traffic load, a higher redundancy level can be beneficial because additional packet copies improve the chance of timely delivery.

However, traditional deterministic multi-path replication provides only coarse-grained redundancy control.
The number of selected paths is discrete, and the number of transmissions of each packet is always an integer.
For example, a packet may be sent once, twice, or three times, but the redundancy level cannot be smoothly adjusted between these integer choices.
Such discrete control may be too coarse for vehicular networks, where link quality, mobility, and traffic load change over time.

To this end, a novel routing scheme termed \emph{Redundancy-Controlled Stochastic Routing (RATIO)} is proposed in this paper.
The main idea of RATIO is simple: instead of forcing each flow to use a single path or blindly replicating every packet over multiple paths, RATIO lets each flow use a compact set of forwarding opportunities and controls how frequently each forwarding link is used.
Specifically, each flow is represented by a reduced directed acyclic graph (DAG), and forwarding probabilities are assigned to the outgoing links in this DAG.
These probabilities determine both the preferred forwarding directions and the average amount of redundancy introduced during packet delivery.
For instance, some packets may be forwarded through one outgoing link while a certain fraction of packets are duplicated over two outgoing links, leading to an average redundancy level such as 1.3 or 1.7.
In this way, RATIO enables continuous tuning of redundancy and provides more precise control than conventional multi-path replication, allowing the routing scheme to better adapt to changing network dynamics.
It should be note that ``continuous'' tuning of redundancy means that the expected number of transmissions for each packet can be any positive value no smaller than one, although the actual transmissions of each individual packet are still discrete.

The ideal RATIO design can be formulated as an optimization problem that jointly determines the reduced DAG and the forwarding probabilities.
The objective is to reduce transmission load while satisfying link-capacity constraints, delay budgets, and reliability requirements of different flows.
However, this problem is difficult to solve exactly for two main reasons.
First, the end-to-end delay and timely packet delivery ratio (PDR) under stochastic forwarding and time-varying vehicular connectivity are not analytically tractable in closed form.
Second, the construction of the reduced DAG itself is combinatorial: different subsets of links and paths may lead to different reliability, delay, and load trade-offs.
This combinatorial structure makes the exact RATIO optimization NP-hard and unsuitable for real-time deployment in large-scale vehicular networks.

Therefore, a practical heuristic scheme, termed \emph{H-RATIO} is further developed in this paper. 
H-RATIO follows the same principle as RATIO but avoids solving the exact optimization problem.
At a high level, it first identifies a compact and reliable forwarding structure from the current network snapshot, and then adjusts the forwarding probabilities according to lightweight estimates of delivery reliability and transmission load.
If the estimated reliability is insufficient, H-RATIO increases redundancy by allowing more packet copies to be created on promising forwarding links.
If the estimated channel load becomes too high, it reduces redundancy to avoid congestion.
Thus, H-RATIO provides a scalable and deployable realization of the RATIO idea using only short-term network observations and local link indicators.

Finally, extensive trace-driven SUMO/ns-3 co-simulations are conducted to evaluate the proposed scheme.
Urban vehicle mobility traces are generated by SUMO and replayed in ns-3, where packet-level multi-hop transmissions, wireless contention, and queueing effects are simulated.
The proposed H-RATIO scheme is compared with representative baseline routing schemes under different traffic loads and mobility conditions.
The results show that H-RATIO improves timely PDR and delivery efficiency by introducing only necessary redundancy, while avoiding the excessive transmission overhead and congestion effects commonly caused by fixed multi-path replication.

The contributions of this paper are summarized as follows:
\begin{itemize}[]
	\item
	A routing method termed redundancy-controlled stochastic routing (RATIO) is proposed and formally formulated, where per-edge
	forwarding probabilities are optimized over flow-specific reduced DAGs by minimizing total transmission load under link capacity, flow PDR, and flow delay constraints.
	
	\item
	An ideal RATIO design is formulated as a load-minimization problem under reliability and delay constraints, and a practical per-period heuristic RATIO (H-RATIO) is developed to address the non-convexity and intractability of the ideal RATIO optimization under partial and short-term observations,
	where reduced DAGs are constructed using link-quality indicators and forwarding probabilities are assigned via lightweight scoring and iterative replication-factor adjustment.
	
	\item
	Extensive trace-driven SUMO/ns-3 co-simulations are conducted with realistic mobility replay and packet-level forwarding to evaluate the proposed H-RATIO scheme.
	It is shown that, under both moderate and high load, H-RATIO matches or exceeds the timely PDR of fixed replication at much lower transmission load and delivery cost.
\end{itemize}

The rest of this paper is organized as follows.
In Section~\ref{sec:model}, the system model, including the network, traffic,
and routing models, is introduced.
The description and ideal problem formulation of RATIO are given in Section~\ref{sec:srdr-scheme}, and a practical heuristic RATIO (H-RATIO) scheme is developed in Section~\ref{sec:heuristic-srdr}.
In Section~\ref{sec:eval}, the performance of RATIO is evaluated and compared with several baselines using trace-driven mobility and
packet-level simulation.
Finally, the paper is concluded in Section~\ref{sec:conclusion}.

\section{Related Work}
\label{sec:related}

Vehicular multi-hop routing under high mobility has been studied from multiple perspectives, including quality-of-service (QoS)-driven routing, mobility-robust VANET/V2X routing, and redundancy-based forwarding.
The most relevant threads are summarized below.

\subsection{QoS Routing in Mobile Wireless Networks}
\label{subsec:related_qos}

QoS routing aims to satisfy end-to-end constraints such as delay (or deadline), packet delivery ratio (PDR), and bandwidth under time-varying wireless resources~\cite{2001_CommMag_QoS_ad_hoc,2007_CommSurvey_QoS_MANET,2011_ComNet_Routing_survey}.
In multi-hop wireless settings, QoS routing is commonly formulated as constrained shortest-path or multi-constraint path selection, where link weights/constraints reflect latency, loss probability, capacity, or interference.
Such formulations remain computationally challenging even under static graphs, and practical solutions have typically relied on heuristics, relaxations, or decompositions that trade optimality for tractability~\cite{2007_CommSurvey_QoS_MANET,2011_ComNet_Routing_survey}.

Early on-demand QoS routing mechanisms have been proposed for multi-hop mobile networks, where QoS-feasible routes are discovered and maintained on demand under dynamic topology~\cite{2001_INFOCOM_On_demand_QoS}.
A representative direction has been the integration of bandwidth estimation into route selection so that QoS decisions can be made using predicted residual capacity rather than purely hop-count or delay surrogates~\cite{2005_JSAC_QoS_bandwidth_est}.
Anycast and group-oriented variants have also been studied, where mobility awareness has been combined with QoS constraints to improve reachability and service continuity in mobile ad hoc settings~\cite{2015_CompEE_Mobility_QoS_MANET}.
More recently, QoS routing has been revisited in software-defined and cross-layer frameworks, where multiple paths can be jointly optimized under centralized control or cross-layer state, e.g., PHY/MAC indicators~\cite{2020_TWC_SDN_multipath,2024_IoT_QoS_cross_layer}.
QoS-driven routing has also been extended to specialized mobile wireless paradigms, including cognitive radio ad hoc networks, where channel uncertainty and spectrum dynamics further complicate QoS satisfaction~\cite{2021_AdHoc_QoS_Cognitive}, and urban MANETs, where QoS adaptivity is emphasized under dense mobility and heterogeneous traffic~\cite{2025_Sensors_QoS_urban_MANET}.

In vehicular networks, QoS routing has often been coupled with mobility-robust metrics, where stability predictors, e.g., link expiration time (LET), and historical link-quality estimates are incorporated to avoid fragile links while meeting delay/throughput requirements~\cite{2010_PersComm_Stability_QoS,2015_NetCompApp_Stability_routing_survey,2014_JNCA_VANET_survey,2021_ACA_VANET_survey}.
Cross-layer indicators, e.g., channel contention, queueing delay proxies, or estimated link service rates, have also been integrated into routing metrics to improve QoS compliance under shared-medium access~\cite{2014_JNCA_VANET_survey}.
While robustness has been improved compared with purely shortest-path routing, the dominant decision unit has largely remained a \emph{path}, or a small set of paths, that is selected and used deterministically within a routing period.
As a result, QoS violations may still be triggered by transient disruptions, including intermittent blockage, interference spikes, and contention bursts, that cannot be fully captured by averaged metrics, especially when routing updates are constrained by protocol overhead and control latency.

\subsection{VANET/V2X Routing Under Mobility}
\label{subsec:related_mobility}

Classical ad hoc routing protocols, including reactive topology-based designs
(e.g., AODV~\cite{rfc3561_aodv} and DSR~\cite{rfc4728_dsr}) and proactive link-state designs
(e.g., OLSR~\cite{rfc3626_olsr}), have often been observed to degrade in vehicular environments due to frequent
route breaks and the associated route-maintenance overhead under rapid topology changes~\cite{2007_VTMag_routing_survey,2008_ComMag_VANET_tutorial,2014_JNCA_VANET_survey}.
To improve delivery performance, extensive V2X routing research has been developed with an emphasis on geographic routing and stability prediction.

Geographic routing reduces route-maintenance cost by leveraging location information and local neighbor states~\cite{2011_TVT_Intersection_GeoRouting,2012_Globecom_GROOV,2018_VehCom_Geographic_routing,2018_ComNet_PGRP_GeoRouting}.
GPSR is a canonical example that combines greedy forwarding with perimeter recovery~\cite{karp2000gpsr}.
Beyond pure geometry, additional signals, e.g., direction/speed, two-hop neighborhood structure, or trust management, have been incorporated to improve robustness against unreliable neighbors and malicious relays~\cite{2023_AdHoc_Trust_GeoRouting}.
However, geographic routing highly relies on accurate position information and timely neighbor-state updates. In highly dynamic vehicular environments, GPS errors, outdated local information, and intermittent NLOS blockage may lead to unreliable next-hop selection and frequent recovery operations.

Complementary to geographic designs, stability-oriented metrics such as link-expiration-time (LET) predictors have been widely used to prefer longer-lived links and paths under mobility~\cite{2010_PersComm_Stability_QoS,2015_NetCompApp_Stability_routing_survey,2019_TVT_Stability_SDN_VehNet,2025_AdHoc_Stability_VANET}.
Future connectivity has typically been estimated from historical link-quality observations and mobility patterns~\cite{2015_NetCompApp_Stability_routing_survey}, and ML-based link-stability prediction has also been explored to improve decision making under uncertainty~\cite{2018_Globelcom_ML_route}.
Nevertheless, the effectiveness of stability-driven routing remains bounded by prediction accuracy. Even if a path is predicted to be relatively stable, it may still be disrupted by unexpected NLOS blockage, interference spikes, or contention bursts. Moreover, most stability-driven methods still make routing decisions at the path level and do not explicitly control how much redundancy should be introduced when stability prediction is uncertain. In RATIO, stability indicators are used as filters or priors during reduced-DAG construction, while controlled redundancy is further leveraged to provide additional resilience against short-term disruptions.

\subsection{Opportunistic Routing and Multipath Routing}
\label{subsec:related_or_multipath}

Opportunistic routing exploits the broadcast nature of wireless channels by selecting forwarders dynamically among candidates that successfully receive a packet~\cite{2022_IICETA_Opportunistic_routing}.
Related ideas also appear in store-carry-forward mechanisms and epidemic-style spreading, which can improve delivery probability at the cost of increased delay and transmission overhead, and are therefore particularly suitable for delay-tolerant regimes~\cite{vahdat2000epidemic}.
ExOR coordinates such forwarding choices to improve throughput and reliability~\cite{biswas2005exor}, and network-coding-assisted variants such as MORE combine opportunistic reception with coded transmissions to reduce coordination overhead and improve robustness~\cite{chachulski2007more}.
A broad overview of opportunistic routing and its design space is provided in~\cite{chakchouk2015or_survey}.
While opportunistic designs can harvest receiver diversity, uncontrolled replication, or overly aggressive forwarder sets, can substantially increase channel contention in V2X, particularly under hidden-terminal effects and rapidly varying link conditions~\cite{chakchouk2015or_survey}.
Thus, opportunistic routing improves robustness by exploiting diversity, but its reliability gain may come with uncontrolled or hard-to-predict transmission overhead. This becomes problematic for low-latency vehicular services, where excessive contention and queueing can reduce timely PDR.

Multipath routing improves reliability by exploiting path diversity and providing alternate delivery attempts when a primary path fails.
Representative designs extend on-demand routing to maintain multiple next hops/paths, e.g., via on-demand multipath distance-vector mechanisms~\cite{marina2001ondemand_multipath}.
In VANET/V2X settings, multipath routing has been widely studied for bandwidth-intensive traffic such as video streaming, often coupled with optimization-based path selection and rate allocation to improve throughput and reduce packet loss~\cite{2018_GCWCN_Multiph_video,2020_PerCom_Optimal_multipath,2021_VTC_Multipath_video}.
These methods typically select a small set of paths and then deterministically split or replicate traffic among them.
However, in highly loaded V2X networks, aggressive multipath usage can inflate channel occupancy and worsen collision-induced losses and latency, causing reliability degradation once the network enters a congested regime~\cite{2014_JNCA_VANET_survey}.
More importantly, traditional multipath replication provides only coarse-grained redundancy control because the number of paths and the number of packet copies are discrete.
The proposed RATIO routing scheme addresses this issue by enabling fine-grained redundancy control beyond integer path replication.

In summary, existing studies provide important tools for vehicular multi-hop delivery, including QoS-aware path selection, mobility-aware/stability-based routing, opportunistic forwarding, and multipath routing. However, a common limitation is that redundancy is either not explicitly controlled or is controlled only in a coarse-grained manner. QoS and stability-based routing mainly improve path selection but remain vulnerable to transient link disruptions. Opportunistic and multipath routing improve reliability through diversity, but may introduce excessive channel occupancy, contention, and delay when redundancy is not carefully regulated. These limitations motivate the design of RATIO, which treats redundancy as a controllable routing resource. By assigning forwarding probabilities over a reduced DAG, RATIO can tune the expected redundancy level continuously according to mobility, traffic load, and per-flow reliability requirements, thereby balancing timely delivery reliability and transmission overhead.

\section{System Model}
\label{sec:model}

In this section, the system model of the considered V2X network is presented, including the network model, traffic model, routing model, and performance metrics.

\subsection{Network Model}

Consider a vehicular network composed of core server, several roadside units (RSUs), and multiple vehicles.
A time-varying directed graph is used to capture connectivity among these nodes.
Specifically, the network at time $t$ is represented by
\begin{equation*}
	G(t) = \bigl(\mathcal{V}, \mathcal{E}(t)\bigr),
\end{equation*}
where $\mathcal{V}$ is the node set and $\mathcal{E}(t)$ is the time-varying set of directed links.
The node set is partitioned as
\begin{equation*}
	\mathcal{V} = \{v_{\text{core}}\} \cup \mathcal{V}_{\text{rsu}} \cup \mathcal{V}_{\text{veh}},
\end{equation*}
where $v_{\text{core}}$ denotes a core server (CORE), $\mathcal{V}_{\text{rsu}}$ denotes the set of roadside units (RSUs), and $\mathcal{V}_{\text{veh}}$ denotes the set of vehicular nodes.

Each vehicle $i \in \mathcal{V}_{\text{veh}}$ is associated with a mobility trajectory $\mathbf{p}_i(t)\in\mathbb{R}^2$ (or $\mathbb{R}^3$) over time.
A directed wireless link $(i,j)$ is considered available at time $t$ when link connectivity conditions (e.g., the received signal strength exceeds a threshold) are satisfied.
The resulting wireless link set is denoted by $\mathcal{E}_{\text{wireless}}(t)$.
In addition, wired backhaul links (usually high-capacity) are included to represent CORE--RSU connections.
These links are collected in $\mathcal{E}_{\text{wired}}$ and are assumed to be highly reliable and always available.
Accordingly, the overall link set is written as
\begin{equation*}
	\mathcal{E}(t) = \mathcal{E}_{\text{wired}} \cup \mathcal{E}_{\text{wireless}}(t).
\end{equation*}

\subsection{Traffic Model}

Let $\mathcal{F}$ denote the set of unicast flows.
Each flow $f\in\mathcal{F}$ is described by
\begin{equation*}
	f = \bigl( s_f, d_f, r_f, t_f^{\text{start}}, t_f^{\text{end}} \bigr),
\end{equation*}
where $s_f \in \mathcal{V}$ is the source node, $d_f \in \mathcal{V}_{\text{veh}}$ is the destination vehicle, $r_f$ is the application bit rate, and $[t_f^{\text{start}}, t_f^{\text{end}})$ is the active time interval of the flow.
We assume a fixed packet size of $L_{\text{pkt}}$ bits, and hence a constant-rate flow with bit rate $r_f$ generates packets at mean rate $r_f/L_{\text{pkt}}$ packets per second during its active interval.

Typical QoS requirements are imposed on each flow, including (i) a maximum end-to-end delay budget $D_f^{\max}$, and (ii) a minimum timely delivery requirement expressed via packet delivery ratio (PDR) target $\eta_f$.

\subsection{Routing Model}

A centralized routing controller is assumed to operate over routing periods of fixed duration $T>0$.
Routing periods are indexed by
\begin{equation*}
	\mathcal{K} = \{0,1,2,\dotsc\},
\end{equation*}
where period $k\in\mathcal{K}$ corresponds to the time interval $[kT,(k+1)T)$.
Let $a_f^{(k)}$ and $\mathcal{F}^{(k)}$ denote the the flow activity and active-flow set in period $k$, respectively.
We have 
\begin{equation*}
	a_f^{(k)} =
	\begin{cases}
		1, & \text{if } kT < t_f^{\text{end}} \text{ and } (k+1)T > t_f^{\text{start}}, \\[1ex]
		0, & \text{otherwise,}
	\end{cases}
\end{equation*}
and
\begin{equation*}
	\mathcal{F}^{(k)} = \bigl\{ f \in \mathcal{F} : a_f^{(k)} = 1 \bigr\}.
\end{equation*}

At the beginning of each period (i.e., at $t=kT$), a connectivity snapshot $G^{(k)}$ is observed:
\begin{equation*}
	G^{(k)} = \bigl(\mathcal{V}, \mathcal{E}^{(k)}\bigr), \qquad
	\mathcal{E}^{(k)} = \mathcal{E}(kT).
\end{equation*}
Based on $G^{(k)}$ and short-term link estimates, routing decisions are computed and then applied throughout $[kT,(k+1)T)$.

For each active flow $f\in\mathcal{F}^{(k)}$, a directed forwarding subgraph $\mathcal{D}_f^{(k)}$ is induced:
\begin{equation*}
	\mathcal{D}_f^{(k)} = \bigl( \mathcal{V}_f^{(k)}, \mathcal{E}_f^{(k)} \bigr),
\end{equation*}
where $\mathcal{E}_f^{(k)} \subseteq \mathcal{E}^{(k)}$ collects the directed edges that may carry packets of flow $f$ during period $k$, and $\mathcal{V}_f^{(k)}$ collects the participating relay/receiver nodes.
For deterministic single-path routing, $\mathcal{D}_f^{(k)}$ reduces to a directed path from $s_f$ to $d_f$.

At the node level, routing decisions are specified as follows.
For each flow $f$, period $k$, node $i$, and outgoing neighbor $j$:
\begin{itemize}
	\item In deterministic single-path schemes, a binary decision $x_{ij}^{f,(k)} \in \{0,1\}$ is used to indicate whether packets of flow $f$ are forwarded from $i$ to $j$ in period $k$.
	\item In probabilistic or load-splitting schemes, a forwarding weight (or probability) $\pi_{ij}^{f,(k)} \in [0,1]$ is assigned to indicate the fraction or probability with which packets are forwarded along $(i,j)$ in period $k$.
\end{itemize}
Different routing strategies are distinguished by how $\mathcal{D}_f^{(k)}$ is constructed and how $\{x_{ij}^{f,(k)}\}$ or $\{\pi_{ij}^{f,(k)}\}$ is assigned.

\subsection{Performance Metrics}
\label{subsec:metrics}

To characterize the reliability--cost trade-off of routing schemes, the timely packet delivery ratio (PDR), also termed as the delay-constrained PDR, is used as the reliability metric, while the total transmission load over all links is used to quantify transmission cost.

Specifically, for a given flow $f$, let $N_f$ denote the number of unique packets generated at the source over the evaluation horizon.
Let $N_f^{\mathrm{del}}(D_f^{\max})$ denote the number of unique packets that are successfully received by the destination $d_f$ within the deadline $D_f^{\max}$ (duplicate receptions are excluded).
The timely PDR of flow $f$ is defined as
\begin{equation*}
	\mathrm{PDR}_f = \frac{N_f^{\mathrm{del}}(D_f^{\max})}{N_f}.
\end{equation*}

To measure transmission cost, for each routing period $k$, let $L_{ij}^{(k)}$ denote the transmission load on directed link $(i,j)$ during period $k$ (including transmissions of replicated copies).
The aggregate load in period $k$, denoted by $L^{(k)}$, is then defined as
\begin{equation*}
	L^{(k)} = \sum_{(i,j)\in \mathcal{E}^{(k)}} L_{ij}^{(k)},
\end{equation*}
and the total transmission load over the evaluation horizon, denoted by $L_{\text{tot}}$, is given by
\begin{equation*}
	L_{\text{tot}} = \sum_{k} L^{(k)}.
\end{equation*}
This metric naturally captures the cost of replication: increasing the replication level for a flow increases $\{L_{ij}^{(k)}\}$ and thus consumes more network resources.

\section{Redundancy-Controlled Stochastic Routing (RATIO) Scheme}
\label{sec:srdr-scheme}

In this section, the proposed Redundancy-Controlled Stochastic Routing (RATIO) scheme is described, followed by an idealized formulation that clarifies the underlying design trade-offs.

\begin{figure}[t]
	\centering
	\includegraphics[width=0.8\linewidth]{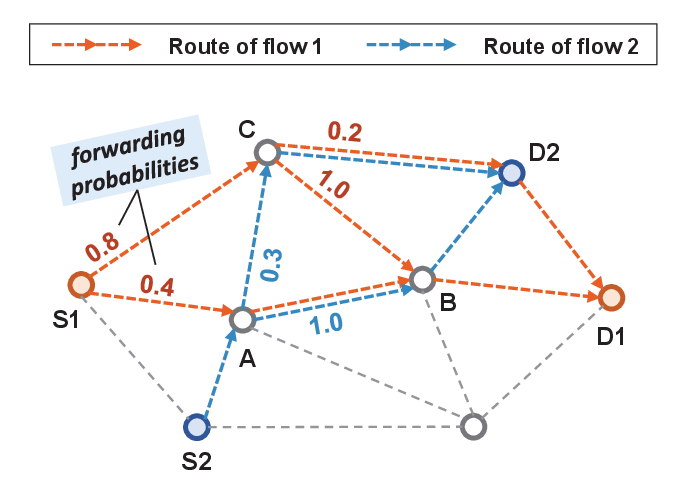}
	\caption{Illustration of the proposed redundancy-controlled stochastic routing (RATIO) scheme.}
	\label{fig:RATIO_scheme}
	\vspace{-12pt}
\end{figure}

\begin{figure}[t]
	\centering
	\resizebox{\columnwidth}{!}{
		\begin{tikzpicture}[scale=1.0, transform shape, font=\small, line cap=round, line join=round]
			
			\definecolor{edgeA}{RGB}{31,119,180}  
			\definecolor{edgeB}{RGB}{255,127,14}  
			\definecolor{edgeC}{RGB}{44,160,44}   
			\definecolor{axisgray}{RGB}{80,80,80}
			
			\def\barH{0.35}
			\def\gapY{0.10}
			\def\rowSep{0.48}
			\def\panelSep{2.35}
			\def\panelBShift{0.4}
			
			\def\UtextDxA{0.22}
			\def\UtextDxB{0.28}
			\def\UtextDy{-0.15}
			
			\def\barsShiftA{0.15}
			\def\barsShiftB{-0.1}
			
			\newcommand{\DrawAxis}[3]{%
				\draw[->,axisgray] (0,#1) -- (1.05,#1) node[right] {#2};
				\foreach \t/\lab in #3 {
					\draw[axisgray] (\t,#1-0.03) -- (\t,#1+0.03);
					\node[axisgray,below] at (\t,#1-0.03) {\footnotesize \lab};
				}
			}
			
			\begin{scope}[x=8.0cm, y=1.0cm, shift={(0,0)}]
				\node[anchor=west] at (0,1.20) {\textbf{(a)} No replication: $j_1=0.6$, $j_2=0.4$ (sum to 1)};
				
				\begin{scope}[yshift=\barsShiftA cm]
					\DrawAxis{0.10}{}{{0/0,0.6/0.6,1/1}}
					
					\path[fill=edgeA, fill opacity=0.45, draw=edgeA]
					(0,0.10+\gapY) rectangle (0.6,0.10+\gapY+\barH);
					\node[axisgray] at (0.30,0.10+\gapY+\barH/2) {\small $\boldsymbol{j_1}$};
					
					\path[fill=edgeB, fill opacity=0.45, draw=edgeB]
					(0.6,0.10+\gapY) rectangle (1.0,0.10+\gapY+\barH);
					\node[axisgray] at (0.80,0.10+\gapY+\barH/2) {\small $\boldsymbol{j_2}$};
					
					\def\Ua{0.64}
					\draw[very thick] (\Ua,\gapY-0.2) -- (\Ua,0.10+\gapY+\barH+0.15);
					\node[below] at (\Ua,\gapY-\barH+\UtextDy) {\small $U=0.64$};
					\node[below] at (\Ua+\UtextDxA,\gapY-\barH+\UtextDy) {(forward to: $j_2$)};
				\end{scope}	
			\end{scope}
			
			\begin{scope}[x=8.0cm, y=1.0cm, shift={(0,-\panelSep-\panelBShift)}]
				\node[anchor=west] at (0,1.45) {\textbf{(b)} Replication possible: $j_1=0.6$, $j_2=0.6$, $j_3=0.4$ (sum to 1.6)};
				
				\begin{scope}[yshift=\barsShiftB cm]
					\DrawAxis{0.10}{}{{0/0,0.2/0.2,0.6/0.6,1/1}}
					
					\draw[axisgray, densely dotted] (0.2,0.10-0.10) -- (0.2,0.10+0.10+2*\rowSep);
					\draw[axisgray, densely dotted] (0.6,0.10-0.10) -- (0.6,0.10+0.10+2*\rowSep);
					
					\def\yRowOne{0.10+\gapY+\rowSep}
					
					\path[fill=edgeA, fill opacity=0.45, draw=edgeA]
					(0,\yRowOne) rectangle (0.6,\yRowOne+\barH);
					\node[axisgray] at (0.30,\yRowOne+\barH/2) {\small $\boldsymbol{j_1}$};
					
					\path[fill=edgeB, fill opacity=0.45, draw=edgeB]
					(0.6,\yRowOne) rectangle (1.0,\yRowOne+\barH);
					\node[axisgray] at (0.80,\yRowOne+\barH/2) {\small $\boldsymbol{j_2}$};
					
					\def\yRowTwo{0.10+\gapY}
					
					\path[fill=edgeB, fill opacity=0.45, draw=edgeB]
					(0.0,\yRowTwo) rectangle (0.2,\yRowTwo+\barH);
					\node[axisgray] at (0.10,\yRowTwo+\barH/2) {\small $\boldsymbol{j_2}$};
					
					\path[fill=edgeC, fill opacity=0.45, draw=edgeC]
					(0.2,\yRowTwo) rectangle (0.6,\yRowTwo+\barH);
					\node[axisgray] at (0.40,\yRowTwo+\barH/2) {\small $\boldsymbol{j_3}$};
					
					\def\Ub{0.35} 
					\draw[very thick] (\Ub,-0.1) -- (\Ub,\yRowOne+\barH+0.15);
					\node[below] at (\Ub,\gapY-\barH+\UtextDy) {\small $U=0.35$};
					\node[below] at (\Ub+\UtextDxB,\gapY-\barH+\UtextDy) {(forward to: $j_1$ and $j_3$)};
				\end{scope}
			\end{scope}
			
		\end{tikzpicture}
	}
	\caption{Illustration of the modulo-based packet forwarding rule in RATIO using a single random draw $U\in[0,1)$. (a) No replication: each $U$ selects exactly one outgoing link.
		(b) Replication enabled: a single $U$ may activate two outgoing links.
		Specifically, $U\in[0,0.2)$ activates $(j_1,j_2)$, $U\in[0.2,0.6)$ activates $(j_1,j_3)$, and $U\in[0.6,1)$ activates $j_2$ only.}
	
	\label{fig:modulo-map}
	\vspace{-12pt}
\end{figure}
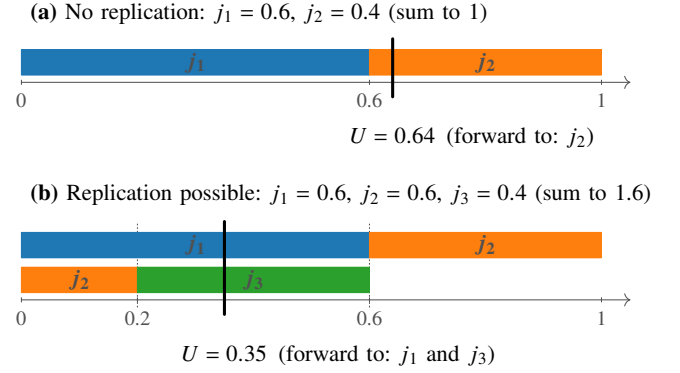

\subsection{RATIO Overview}
\label{subsec:srdr-scheme-overview}

An illustration of RATIO is provided in Fig.~\ref{fig:RATIO_scheme}.
The main idea of RATIO is to control redundancy in a gradual manner rather than making a fixed single-path or multi-path replication decision.
For each flow, RATIO first restricts the routing space to a compact reduced DAG, which contains several promising forwarding opportunities from the source to the destination.
Then, instead of selecting only one path or replicating every packet over all available paths, RATIO assigns forwarding probabilities to the outgoing links in the reduced DAG.
These probabilities determine how often each link is used and how much redundancy is introduced on average.

At a non-fork node with only one outgoing link, packets are forwarded deterministically along that link.
At a fork node with multiple outgoing links, an incoming packet can be forwarded to one or more next-hop nodes according to the assigned forwarding probabilities.
If the aggregate forwarding probability at a fork node is close to one, the node behaves similarly to probabilistic path splitting and usually generates one outgoing copy per packet.
If the aggregate forwarding probability is larger than one, additional packet copies can be generated with controlled frequency.
In this way, RATIO can increase redundancy when link uncertainty is high and reduce redundancy when the network is congested.

A key point is that RATIO controls redundancy in expectation.
Although each individual packet transmission is still discrete, the expected number of outgoing copies can be continuously tuned by adjusting the forwarding probabilities.
For example, a fork node may generate on average $1.5$ outgoing copies per incoming packet: some packets are forwarded on one outgoing link, while others are replicated over two links.
This provides finer-grained control than conventional deterministic multi-path replication, where the number of packet copies is usually an integer.

The packet-level forwarding decision is implemented by a modulo-based stochastic forwarding rule as illustrated in Fig.~\ref{fig:modulo-map}.
This rule maps the continuous-valued forwarding probabilities to actual discrete packet transmissions, while ensuring that at least one outgoing link is selected whenever forwarding is feasible.
The detailed components of RATIO are presented below.

\subsection{Components of RATIO}

The key components of RATIO is reduced DAG as routing structure and forwarding probabilities that guide packet transmission and duplication.
The overall node operations of RATIO at each routing period are summarized in Algorithm~\ref{alg:srdr-forwarding}.

\begin{algorithm}[t]
	\caption{RATIO forwarding at node $i$ for flow $f$ in period $k$}
	\label{alg:srdr-forwarding}
	\begin{algorithmic}[1]
		\REQUIRE Reduced DAG $\mathcal{D}_f^{(k)}$; outgoing set $\mathcal{N}_{f,i}^{+,(k)}$; forwarding probs $\{\pi_{ij}^{f,(k)}\}$
		\ENSURE A set of outgoing transmissions (possibly multiple)
		
		\IF{$i=d_f$}
		\STATE Deliver to upper layer.
		\ELSIF{$\mathcal{N}_{f,i}^{+,(k)}=\emptyset$}
		\STATE Drop packet.
		\ELSIF{$|\mathcal{N}_{f,i}^{+,(k)}|=1$}
		\STATE Transmit on $(i,j)$.
		\ELSE
		\STATE (Randomly) order $\mathcal{N}_{f,i}^{+,(k)}=\{j_1,\dots,j_m\}$.
		\STATE $s_0\gets 0$, $s_\ell\gets \sum_{u=1}^{\ell}\pi_{ij_u}^{f,(k)}$ for $\ell=1,\dots,m$.
		\STATE Draw $U\sim\mathrm{Unif}(0,1)$.
		\FOR{$\ell=1$ to $m$}
		\STATE $a\gets s_{\ell-1}$, $b\gets s_\ell$.
		\STATE $a_0\gets a-\lfloor a\rfloor$, $b_0\gets b-\lfloor b\rfloor$.
		\IF{$\lfloor a\rfloor=\lfloor b\rfloor$}
		\IF{$U\in[a_0,b_0)$}
		\STATE Transmit a copy on $(i,j_\ell)$.
		\ENDIF
		\ELSE
		\IF{$U\in[a_0,1)$ \textbf{or} $U\in[0,b_0)$}
		\STATE Transmit a copy on $(i,j_\ell)$.
		\ENDIF
		\ENDIF
		\ENDFOR
		\ENDIF
	\end{algorithmic}
\end{algorithm}

\subsubsection{Reduced DAG}

The reduced DAG can be interpreted as the candidate forwarding region of a flow.
It is not intended to include all possible links in the network.
Instead, it keeps only a limited set of useful forwarding links so that the routing decision remains tractable while still preserving path diversity.
Compared with a single path, the reduced DAG provides alternative forwarding opportunities under mobility.
Compared with using the full network graph, it avoids excessive search complexity and unnecessary redundancy.

In routing period $k$, network connectivity is represented by a directed graph $G^{(k)}=(\mathcal{V},\mathcal{E}^{(k)})$.
For each active flow $f$ in period $k$, a flow-specific reduced DAG $\mathcal{D}_f^{(k)}$ that connects $s_f$ to $d_f$ is utilized as the route space:
\begin{equation*}
	\mathcal{D}_f^{(k)}=\bigl(\mathcal{V}_f^{(k)},\mathcal{E}_f^{(k)}\bigr), \qquad
	\mathcal{V}_f^{(k)}\subseteq\mathcal{V},~ \mathcal{E}_f^{(k)}\subseteq\mathcal{E}^{(k)}.
\end{equation*}
The reduced DAG is constructed to keep the routing search space compact while preserving a limited amount of path diversity. 

For each node $i\in\mathcal{V}_f^{(k)}$, the outgoing and incoming neighbor sets in $\mathcal{D}_f^{(k)}$, denoted by $\mathcal{N}_{f,i}^{+,(k)}$ and $\mathcal{N}_{f,i}^{-,(k)}$, are defined as
\begin{align*}
	\mathcal{N}_{f,i}^{+,(k)} &= \{j\in\mathcal{V}_f^{(k)}:(i,j)\in\mathcal{E}_f^{(k)}\},\\
	\mathcal{N}_{f,i}^{-,(k)} &= \{j\in\mathcal{V}_f^{(k)}:(j,i)\in\mathcal{E}_f^{(k)}\},
\end{align*}
respectively.
A node with $|\mathcal{N}_{f,i}^{+,(k)}|\ge 2$ is referred to as a fork node.

\subsubsection{Forwarding Probabilities and Replication Factor}

Each directed edge $(i,j)\in\mathcal{E}_f^{(k)}$ is associated with a forwarding probability
\begin{equation*}
	\pi_{ij}^{f,(k)}\in[0,1],
\end{equation*}
which is interpreted as the marginal probability that an arriving packet of flow $f$ at node $i$ is forwarded on link $(i,j)$ during period $k$.
Intuitively, $\pi_{ij}^{f,(k)}$ controls how frequently link $(i,j)$ is used by flow $f$ at node $i$.
A larger $\pi_{ij}^{f,(k)}$ means that packets are more likely to be forwarded through link $(i,j)$, while $\pi_{ij}^{f,(k)}=0$ means that this link is not used by the flow.
Therefore, forwarding probabilities jointly determine both the preferred forwarding direction and the amount of traffic injected onto each outgoing link.

The aggregate replication factor at node $i$, denoted by $\Pi_i^{f,(k)}$, is then given by
\begin{equation*}
	\Pi_i^{f,(k)} = \sum_{j\in\mathcal{N}_{f,i}^{+,(k)}} \pi_{ij}^{f,(k)},
\end{equation*}
which represents the expected number of outgoing copies generated per incoming packet at node $i$.
Unlike conventional probabilistic splitting (which typically enforces $\Pi_i^{f,(k)}=1$), values $\Pi_i^{f,(k)}>1$ are permitted at fork nodes so that replication can be introduced in a continuous and tunable manner.
The aggregate replication factor $\Pi_i^{f,(k)}$ represents the expected number of outgoing packet copies generated by node $i$ for each incoming packet of flow $f$.
When $\Pi_i^{f,(k)}=1$, the node forwards each packet along one outgoing direction on average, which corresponds to no additional replication.
When $\Pi_i^{f,(k)}>1$, the node creates additional copies with a controlled frequency.
For example, $\Pi_i^{f,(k)}=1.5$ means that, over many packets, the node generates $1.5$ outgoing copies per incoming packet on average.
This does not mean that each individual packet is transmitted $1.5$ times; rather, some packets may be forwarded once, while others may be replicated over multiple outgoing links.
Thus, RATIO provides continuous-valued redundancy control in expectation, while packet-level transmissions remain discrete.

To avoid intentional packet dropping at relay nodes, the following consistency condition is imposed:
\begin{equation}
	\begin{aligned}
		\sum_{j\in\mathcal{N}_{f,i}^{+,(k)}} \pi_{ij}^{f,(k)}
		&\ge 1, \\
		&\forall f,~\forall k,~\forall i\in\mathcal{V}_f^{(k)},~
		\mathcal{N}_{f,i}^{+,(k)}\neq\emptyset.
	\end{aligned}
	\label{constr:node-forward}
\end{equation}
In particular, when $|\mathcal{N}_{f,i}^{+,(k)}|=1$, \eqref{constr:node-forward} implies $\pi_{ij}^{f,(k)}=1$, i.e., non-fork nodes deterministically forward along their unique outgoing edge.

It should be noted that this continuity is defined at the level of expected redundancy rather than the level of physical packet transmissions. The forwarding probabilities $\{\pi_{ij}^{f,(k)}\}$ are continuous-valued variables, and their sum $\Pi_i^{f,(k)}$ represents the expected number of outgoing copies generated per incoming packet. For instance, $\Pi_i^{f,(k)}=1.5$ does not imply that every packet is literally transmitted $1.5$ times. Instead, each packet-level forwarding decision remains discrete: a packet copy is either transmitted or not transmitted on each outgoing link. The modulo-based stochastic forwarding rule maps the continuous-valued probabilities to discrete packet transmissions, so some packets may generate one outgoing copy while others may generate two or more copies. Over a sequence of packets, the average number of outgoing copies approaches the prescribed value $\Pi_i^{f,(k)}$. Therefore, RATIO provides continuous-valued redundancy control in expectation, while the realized redundancy is stochastic across packets.

\subsubsection{Packet-Level Forwarding Rule at Nodes}

The marginal forwarding probabilities are realized using a single random draw per packet (or packet copy) at each fork node, while ensuring that at least one outgoing link is selected via a modulo-based mapping.
An illustration of the forwarding rule is shown in Fig.~\ref{fig:modulo-map}.

To be specific, consider a fork node $i$ with ordered outgoing neighbors $\mathcal{N}_{f,i}^{+,(k)}=\{j_1,\dots,j_m\}$, where $m\ge 2$.
Define cumulative sums
\begin{equation*}
	s_0 = 0,\qquad
	s_\ell = \sum_{u=1}^{\ell}\pi_{ij_u}^{f,(k)},~~\ell=1,\dots,m. 
\end{equation*}
Each edge $(i,j_\ell)$ is assigned an interval
\begin{equation*}
	I_{ij_\ell}^{f,(k)} = [s_{\ell-1}, s_\ell) \subseteq [0,\Pi_i^{f,(k)}) .
\end{equation*}
Let $\varphi(x)= x \bmod 1 \in [0,1)$ denote the modulo map.
For each arriving packet, a single random variable $U\sim\mathrm{Unif}(0,1)$ is drawn, and a copy is forwarded on every edge $(i,j_\ell)$ such that
\begin{equation*}
	U \in \varphi\!\left(I_{ij_\ell}^{f,(k)}\right).
\end{equation*}
This modulo-based rule converts continuous forwarding probabilities into actual packet-level decisions.
The same random number $U$ is used to test all outgoing intervals.
If $U$ falls into the mapped interval of one outgoing link, a copy is transmitted on that link.
If $U$ falls into the mapped intervals of multiple outgoing links, multiple copies are transmitted.
Therefore, a packet may be sent through one link or replicated across several links, depending on the random draw and the forwarding probabilities.

When $\Pi_i^{f,(k)}\ge 1$, the mapped intervals $\{\varphi(I_{ij_\ell}^{f,(k)})\}_{\ell=1}^m$ jointly cover $[0,1)$, ensuring that at least one outgoing transmission is triggered whenever forwarding is feasible.
More specifically, with ordinary independent probabilistic forwarding, it is possible that all outgoing links reject the packet, which would create unintended packet loss.
In contrast, when $\Pi_i^{f,(k)}\ge 1$, the modulo-mapped intervals cover the whole interval $[0,1)$, so at least one outgoing link is activated.
At the same time, overlapping mapped intervals naturally realize controlled replication.

\subsection{Problem Formulation}
\label{subsec:problem}

In each routing period $k$, the RATIO design is composed of two coupled components:
(i) a reduced DAG $\mathcal{D}_f^{(k)}$ is selected for each active flow $f$, and
(ii) forwarding probabilities $\{\pi_{ij}^{f,(k)}\}$ are assigned on the selected DAG edges.
An idealized joint formulation is provided below.

\subsubsection{Decision Variables}

For each flow $f$ and period $k$, a binary edge-selection variable is introduced to represent the reduced DAG:
\begin{equation*}
	z_{ij}^{f,(k)}\in\{0,1\}, \qquad \forall (i,j)\in\mathcal{E}^{(k)} ,
\end{equation*}
where $z_{ij}^{f,(k)}=1$ indicates that link $(i,j)$ is included in the reduced DAG $\mathcal{D}_f^{(k)}$ for flow $f$.
Forwarding probabilities are defined on physical edges, but become effective only when the edge is selected:
\begin{equation*}
	\pi_{ij}^{f,(k)}\in[0,1], \qquad \forall (i,j)\in\mathcal{E}^{(k)},
\end{equation*}
together with the coupling constraint
\begin{equation}
	0 \le \pi_{ij}^{f,(k)} \le z_{ij}^{f,(k)}, \qquad \forall (i,j)\in\mathcal{E}^{(k)} .
	\label{constr:pi_z}
\end{equation}
Besides, if $(i,j)\notin\mathcal{E}_f^{(k)}$ (i.e., $z_{ij}^{f,(k)}=0$), then $\pi_{ij}^{f,(k)}=0$ is enforced automatically.

The two types of variables have different roles.
The binary variable $z_{ij}^{f,(k)}$ decides whether link $(i,j)$ is included in the reduced DAG of flow $f$.
The continuous variable $\pi_{ij}^{f,(k)}$ further decides how frequently this selected link is used for packet forwarding.
Thus, $z_{ij}^{f,(k)}$ determines the available forwarding structure, while $\pi_{ij}^{f,(k)}$ determines the redundancy level and traffic distribution on that structure.

\subsubsection{Reduced DAG Feasibility Constraints}

To ensure that the selected subgraph is a DAG, integer topological-order variables $u_i^{f,(k)}\in\{0,1,\dots,|\mathcal{V}|-1\}$ are introduced, and the following constraint is imposed:
\begin{equation}
	u_i^{f,(k)} + 1 \le u_j^{f,(k)} + M\bigl(1-z_{ij}^{f,(k)}\bigr), \quad \forall (i,j)\in\mathcal{E}^{(k)}, f, k,
	\label{constr:acyclic_mtz}
\end{equation}
where $M\ge |\mathcal{V}|$ is a sufficiently large constant.
When $z_{ij}^{f,(k)}=1$, equation \eqref{constr:acyclic_mtz} enforces $u_i^{f,(k)}<u_j^{f,(k)}$, thereby precluding directed cycles.

Connectivity of the reduced DAG is enforced by routing one unit of auxiliary flow.
Auxiliary variables $\phi_{ij}^{f,(k)}\in[0,1]$ are introduced and constrained as
\begin{equation}
	\label{constr:unit_flow}
	\begin{aligned}
		& \sum_{j:(i,j)\in\mathcal{E}^{(k)}} \phi_{ij}^{f,(k)}-\sum_{h:(h,i)\in\mathcal{E}^{(k)}} \phi_{hi}^{f,(k)} \\
		&=
		\begin{cases}
			1, & i=s_f,\\
			-1,& i=d_f,\\
			0, & \text{otherwise},
		\end{cases}
		\qquad \forall i\in\mathcal{V},~\forall f,k.
	\end{aligned}
\end{equation}
\begin{equation}
	0 \le \phi_{ij}^{f,(k)} \le z_{ij}^{f,(k)},
	\qquad \forall (i,j)\in\mathcal{E}^{(k)},~\forall f,k.
	\label{constr:flow_couple}
\end{equation}
Constraints \eqref{constr:unit_flow}--\eqref{constr:flow_couple} guarantee the existence of at least one directed $s_f$--$d_f$ path in the selected edges.
Besides, they ensure that packets always make progress along the reduced DAG and cannot circulate in routing loops.

\subsubsection{Forwarding-Consistency Constraints}

To avoid explicitly invoking the outgoing neighbor set $\mathcal{N}_{f,i}^{+,(k)}$ in the optimization (which would otherwise require the reduced DAG structure to be referenced inside the constraints), a binary flag $o_i^{f,(k)}\in\{0,1\}$ is introduced to indicate whether node $i$ has any selected outgoing edge:
\begin{align}
	z_{ij}^{f,(k)} &\le o_i^{f,(k)}, \qquad \forall (i,j)\in\mathcal{E}^{(k)},~\forall f,k,
	\label{constr:o_flag}\\
	\sum_{j:(i,j)\in\mathcal{E}^{(k)}} \pi_{ij}^{f,(k)} &\ge o_i^{f,(k)}, \qquad \forall i\in\mathcal{V},~\forall f,k .
	\label{constr:node_forward_joint}
\end{align}
When $o_i^{f,(k)}=1$, constraint~\eqref{constr:node_forward_joint} enforces $\sum_{j}\pi_{ij}^{f,(k)}\ge 1$, ensuring that at least one outgoing transmission is scheduled whenever node $i$ has selected outgoing edges.
When $o_i^{f,(k)}=0$, no outgoing edge is selected and this constraint becomes vacuous.

\subsubsection{Performance Functions}

Under wireless channel dynamics, medium-access contention, and queueing interactions, end-to-end performance is generally difficult to characterize in closed form.
Accordingly, the timely PDR of flow $f$, denoted by $\mathrm{PDR}_f$, and the total transmission load, denoted by $L_{\mathrm{tot}}$, are treated as (possibly non-convex) functions of the design variables $\{z\}$ and $\{\pi\}$:
\begin{align*}
	\mathrm{PDR}_f &= \mathrm{PDR}_f\bigl(\{z\},\{\pi\}\bigr), \\
	L_{\mathrm{tot}} &= L_{\mathrm{tot}}\bigl(\{z\},\{\pi\}\bigr).
\end{align*}
Here, when computing $\mathrm{PDR}_f$, a packet is counted as successfully delivered if at least one of its copies reaches $d_f$ within the deadline (duplicate receptions are ignored).
But any additional copies beyond the first delivery are still counted in $L_{\mathrm{tot}}$.

\subsubsection{Idealized Formulation}

An idealized RATIO design is formulated by minimizing the total transmission load while satisfying per-flow timely-reliability constraints and per-link capacity constraints:
\begin{subequations}
	\label{prob:RATIO}
	\begin{align}
		& \min_{\{z,\pi,u,\phi,o\}} \quad L_{\mathrm{tot}}\bigl(\{z\},\{\pi\}\bigr) \label{obj:RATIO_joint}\\
		\text{s.t.}~~
		& \mathrm{PDR}_f\bigl(\{z\},\{\pi\};D_{\max}\bigr)\ge \eta_f, ~~ \forall f\in\mathcal{F}, \label{constr:ratio_pdr}\\
		& \sum_{f\in\mathcal{F}^{(k)}} \pi_{ij}^{f,(k)} r_f \le C_{ij}^{(k)},
		~~ \forall k\in\mathcal{K},~(i,j)\in\mathcal{E}^{(k)}, \label{constr:ratio_cap}\\
		& \eqref{constr:pi_z},~\eqref{constr:acyclic_mtz},~
		\eqref{constr:unit_flow}-\eqref{constr:flow_couple},~ \eqref{constr:o_flag}-\eqref{constr:node_forward_joint}, \nonumber \\
		& z_{ij}^{f,(k)}\in\{0,1\},~~ 0\le \pi_{ij}^{f,(k)}\le 1, \nonumber \\
		& u_i^{f,(k)}\in\{0,\dots,|\mathcal{V}|-1\}, \nonumber \\
		& 0\le \phi_{ij}^{f,(k)}\le 1,~~ o_i^{f,(k)}\in\{0,1\}. \nonumber
	\end{align}
\end{subequations}
Here $r_f$ denotes injection rate of flow $f$, and $C_{ij}^{(k)}$ denotes the service capability of link $(i,j)$ in period $k$.

Problem~\eqref{prob:RATIO} characterizes an idealized RATIO design but is difficult to solve exactly.
The difficulty comes from two main aspects.
First, the timely PDR and end-to-end delay under stochastic forwarding, wireless contention, and queueing interactions are not analytically tractable in closed form.
As a result, the functions $\mathrm{PDR}f({z},{\pi})$ and $L{\mathrm{tot}}({z},{\pi})$ cannot be easily optimized by standard convex optimization methods.
Second, the reduced-DAG construction is combinatorial because the controller needs to select a suitable subset of links and paths from the time-varying connectivity graph.
Different DAG choices may lead to different reliability, delay, and load trade-offs.
Therefore, the joint optimization of DAG construction and forwarding probabilities is mixed-integer, generally non-convex, and NP-hard.
In addition, only instantaneous connectivity and short-term MAC/traffic statistics are available in each routing period, which further makes exact horizon-wide optimization impractical for large-scale vehicular networks.

\section{Heuristic Redundancy-Controlled Stochastic Routing (H-RATIO)}
\label{sec:heuristic-srdr}

To enable tractable and scalable RATIO routing, a per-period heuristic, namely H-RATIO, is developed in this section.
Instead of directly solving the idealized RATIO problem, H-RATIO separates the design into two simpler steps:
(i) constructing a compact reduced DAG for each flow, and
(ii) assigning and refining forwarding probabilities on the constructed DAG.
Surrogate models are introduced to estimate timely PDR and transmission load, so that H-RATIO can adjust redundancy using only short-term measurements and lightweight computations.

\subsection{H-RATIO Overview}
H-RATIO is a practical realization of the RATIO idea.
Its goal is not to find the globally optimal reduced DAG and forwarding probabilities, which would be computationally expensive.
Instead, H-RATIO follows a simple and scalable procedure.

First, H-RATIO constructs a reduced DAG for each flow from the current network snapshot.
Links with poor quality or short expected lifetime are removed, and a small number of candidate paths are extracted.
The union of these paths forms a compact forwarding structure that preserves useful path diversity without using the entire network graph.

Second, H-RATIO assigns initial forwarding probabilities on the reduced DAG.
Links with better quality, longer expected lifetime, or shorter remaining distance to the destination are given larger splitting weights.
These weights decide the preferred forwarding directions at each fork node.

Third, H-RATIO adjusts the replication level according to surrogate estimates of timely PDR and link load.
If the estimated timely PDR of a flow is below its target, H-RATIO increases redundancy at selected fork nodes to activate more forwarding opportunities.
If some links become overloaded, H-RATIO decreases redundancy for the flows contributing most to the overload.
Through this iterative adjustment, H-RATIO balances two competing effects: more redundancy can improve reliability, but excessive redundancy may increase congestion and delay.

The overall workflow of H-RATIO is summarized in Algorithm~\ref{alg:heuristic-ratio}.
The surrogate models used for reliability and load estimation are introduced first, followed by the detailed operations of reduced-DAG construction and forwarding-probability refinement.

\begin{algorithm}[t]
	\caption{Heuristic RATIO (H-RATIO) in routing period $k$}
	\label{alg:heuristic-ratio}
	\begin{algorithmic}[1]
		\REQUIRE Connectivity graph $G^{(k)}=(\mathcal{V},\mathcal{E}^{(k)})$; effective rates $\{\bar{C}_{ij}^{(k)}\}$; link indicators $\{\gamma_{ij}^{(k)},\mathrm{LET}_{ij}^{(k)}\}$; active flows $\mathcal{F}^{(k)}$ with $(s_f,d_f,r_f,D_{\max}^f,\eta_f)$; packet size $L_{\mathrm{pkt}}$; thresholds $(\gamma_{\min},\mathrm{LET}_{\min})$; parameters $(P_{\max}, I_{\max}, \Delta, \rho_{\max})$.
		\ENSURE Forwarding probabilities $\{\pi_{ij}^{f,(k)}\}$ on reduced DAGs $\{\mathcal{D}_f^{(k)}\}$.
		
		\STATE Prune links with $\gamma_{ij}^{(k)}<\gamma_{\min}$ or $\mathrm{LET}_{ij}^{(k)}<\mathrm{LET}_{\min}$.
		
		\FOR{each flow $f\in\mathcal{F}^{(k)}$}
		\STATE Construct reduced DAG $\mathcal{D}_f^{(k)}$ from up to $P_{\max}$ candidate paths (Section~\ref{subsec:heuristic-mesh}).
		\STATE Compute local edge scores $\{\psi_{ij}^{f,(k)}\}$ and splitting weights $\{\omega_{ij}^{f,(k)}\}$.
		\STATE Initialize replication factors $\kappa_i^{f,(k)}\leftarrow 1$ and set $\pi_{ij}^{f,(k)} \leftarrow \min\{1,\kappa_i^{f,(k)}\omega_{ij}^{f,(k)}\}$ on $\mathcal{D}_f^{(k)}$.
		\ENDFOR
		
		\FOR{$\mathrm{iter}=1$ to $I_{\max}$}
		\STATE Compute expected node arrival rates $\{x_i^{f,(k)}\}$ on each $\mathcal{D}_f^{(k)}$ via~\eqref{eq:x-recursion-refinement}.
		\STATE Compute link loads $\{\lambda_{ij}^{(k)}\}$ by \eqref{eq:lambda-flow}--\eqref{eq:lambda-ij} and utilizations $\{\rho_{ij}^{(k)}\}$ by \eqref{eq:rho-ij}.
		\FOR{each flow $f\in\mathcal{F}^{(k)}$}
		\STATE Evaluate $\widehat{\mathrm{PDR}}_f^{(k)}(D_{\max}^f)$ via \eqref{eq:path-mean}--\eqref{eq:pdr-flow-union}.
		\IF{$\widehat{\mathrm{PDR}}_f^{(k)}(D_{\max}^f)<\eta_f$}
		\STATE Increase $\kappa_i^{f,(k)}$ by $\Delta$ on selected fork nodes.
		\ENDIF
		\ENDFOR
		\IF{$\exists (i,j)$ with $\rho_{ij}^{(k)}>\rho_{\max}$}
		\STATE Decrease $\kappa_i^{f,(k)}$ on nodes contributing most to overloaded links.
		\ENDIF
		\STATE Update $\pi_{ij}^{f,(k)}$ and $\{x_i^{f,(k)}\}$ and on each reduced DAG via \eqref{eq:pi-heuristic} and \eqref{eq:x-recursion-refinement}, respectively.
		\IF{no $\kappa_i^{f,(k)}$ changes} \STATE \textbf{break} \ENDIF
		\ENDFOR
	\end{algorithmic}
\end{algorithm}

\subsection{Surrogate Models for Timely PDR and Transmission Load}
\label{subsec:heuristic-surrogates}

To enable period-by-period routing decisions with limited information, surrogate models are adopted for both timely PDR and transmission load.
In routing period $k$, an effective service rate $\bar{C}_{ij}^{(k)}$ (bit/s) is assumed to be estimated by the controller for each directed link $(i,j)\in\mathcal{E}^{(k)}$, based on short-term measurements and/or historical statistics.

\subsubsection{Surrogate Transmission Load}

Let $x_i^{f,(k)}$ denote the expected arrival rate of flow $f$ at node $i$ in period $k$, measured in bit/s.
Given forwarding probabilities $\{\pi_{ij}^{f,(k)}\}$, the expected traffic rate of flow $f$ forwarded on link $(i,j)$ is approximated as
\begin{equation}
	\lambda_{ij}^{f,(k)} \triangleq x_i^{f,(k)}\,\pi_{ij}^{f,(k)},
	\qquad (i,j)\in\mathcal{E}_f^{(k)}.
	\label{eq:lambda-flow}
\end{equation}
That is, the traffic placed on link $(i,j)$ is proportional to the forwarding probability on that link.
If $\pi_{ij}^{f,(k)}$ is increased, a larger fraction of the traffic arriving at node $i$ will be forwarded through link $(i,j)$.
Therefore, forwarding probabilities directly determine both the traffic distribution and the redundancy-induced load.

The aggregate offered load on a physical link $(i,j)$ can then be derived by aggregating the traffic contributions of all flows using the same physical link, i.e.,
\begin{equation}
	\lambda_{ij}^{(k)} \triangleq \sum_{f\in\mathcal{F}_{ij}^{(k)}} \lambda_{ij}^{f,(k)},
	\label{eq:lambda-ij}
\end{equation}
where $\mathcal{F}_{ij}^{(k)}$ denotes the set of flows that may use link $(i,j)$ in period $k$.

As a surrogate for transmission cost, the per-period total load is defined as the sum of aggregate link loads:
\begin{equation}
	\overline{L}^{(k)} \triangleq \sum_{(i,j)\in\mathcal{E}^{(k)}} \lambda_{ij}^{(k)} .
	\label{eq:total-load-period}
\end{equation}
The surrogate load $\overline{L}^{(k)}$ therefore measures the overall channel occupation caused by forwarding and replication in routing period $k$.
A smaller $\overline{L}^{(k)}$ indicates lower transmission overhead, while a larger value suggests that more wireless resources are consumed.

Moreover, this surrogate captures the replication effect directly: increasing forwarding probabilities (and hence aggregate replication factors) increases $\{\lambda_{ij}^{(k)}\}$ and therefore increases $\overline{L}^{(k)}$.

\subsubsection{Surrogate Timely PDR}
\label{subsubsec:surrogate_timely_pdr}

To capture contention-induced delay inflation, processing and propagation delays are neglected, while the waiting time plus service time is modeled.
Given the fixed packet size $L_{\mathrm{pkt}}$, the corresponding packet-level arrival and service rates, denoted by $\hat{\lambda}_{ij}^{(k)}$ and $\hat{\mu}_{ij}^{(k)}$, respectively, are given by
\begin{equation*}
	\hat{\lambda}_{ij}^{(k)} = \frac{\lambda_{ij}^{(k)}}{L_{\mathrm{pkt}}} \quad (\text{pkt/s}),
	\qquad
	\hat{\mu}_{ij}^{(k)} =\frac{\bar{C}_{ij}^{(k)}}{L_{\mathrm{pkt}}} \quad (\text{pkt/s}).
\end{equation*}
The utilization of link $(i, j)$, denoted by $\rho_{ij}^{(k)}$, is then given by
\begin{equation}
	\rho_{ij}^{(k)} \triangleq \frac{\hat{\lambda}_{ij}^{(k)}}{\hat{\mu}_{ij}^{(k)}}
	= \frac{\lambda_{ij}^{(k)}}{\bar{C}_{ij}^{(k)}} \in [0,1),
	\label{eq:rho-ij}
\end{equation}
and stability is ensured when $\rho_{ij}^{(k)} < 1$.
The utilization $\rho_{ij}^{(k)}$ indicates how heavily link $(i,j)$ is loaded.
When $\rho_{ij}^{(k)}$ is close to zero, the link has sufficient residual service capability.
When $\rho_{ij}^{(k)}$ approaches one, the link becomes heavily loaded, and queueing delay may increase rapidly.
Thus, $\rho_{ij}^{(k)}$ is used by H-RATIO to detect potential congestion caused by excessive redundancy.

Each link is then approximated by an M/M/1 queue with arrival rate $\hat{\lambda}_{ij}^{(k)}$ and service rate $\hat{\mu}_{ij}^{(k)}$.
Define the (packet-level) service slack as
\begin{equation}
	\beta_{ij}^{(k)} \triangleq \hat{\mu}_{ij}^{(k)}-\hat{\lambda}_{ij}^{(k)}
	= \frac{\bar{C}_{ij}^{(k)}-\lambda_{ij}^{(k)}}{L_{\mathrm{pkt}}} > 0.
	\label{eq:beta-slack}
\end{equation}
The service slack $\beta_{ij}^{(k)}$ measures the remaining service capability after accounting for the offered load.
A larger $\beta_{ij}^{(k)}$ implies that packets can be served more quickly on link $(i,j)$.
A smaller $\beta_{ij}^{(k)}$ means that the link is closer to congestion, which leads to larger waiting time and a lower probability of meeting the deadline.

Under this model, the sojourn time (i.e., waiting time plus service time) on link $(i,j)$, denoted by $T_{ij}^{(k)}$, is exponentially distributed, i.e., 
\begin{equation*}
	T_{ij}^{(k)} \sim \mathrm{Exp}\!\bigl(\beta_{ij}^{(k)}\bigr).
\end{equation*}
Its mean and variance are
\begin{equation*}
	\mathbb{E}\!\left[T_{ij}^{(k)}\right] = \frac{1}{\beta_{ij}^{(k)}}
	= \frac{L_{\mathrm{pkt}}}{\bar{C}_{ij}^{(k)}-\lambda_{ij}^{(k)}},
\end{equation*}
and
\begin{equation*}
	\mathrm{Var}\!\left(T_{ij}^{(k)}\right) = \frac{1}{\bigl(\beta_{ij}^{(k)}\bigr)^2}
	= \frac{L_{\mathrm{pkt}}^2}{\bigl(\bar{C}_{ij}^{(k)}-\lambda_{ij}^{(k)}\bigr)^2},
\end{equation*}
respectively.
The path mean $\mu_{f,p}^{(k)}$ and variance $\bigl(\sigma_{f,p}^{(k)}\bigr)^2$ summarize the delay behavior of path $p$.
If a path contains heavily loaded links, then $\bar{C}{ij}^{(k)}-\lambda{ij}^{(k)}$ becomes small, and both the mean and variance of the path delay increase.
Therefore, a path with many congested links is less likely to deliver packets before the deadline.

For each flow $f$, the reduced DAG $\mathcal{D}_f^{(k)}$ induces a set of candidate $s_f$--$d_f$ paths, denoted by $\mathcal{P}_f^{(k)}$.
Consider any path $p\in\mathcal{P}_f^{(k)}$, consisting of directed edges $e\in p$.
Assuming hop delays are independent across edges, the end-to-end delay along path $p$ is modeled as
\begin{equation*}
	S_{f,p}^{(k)} \triangleq \sum_{e\in p} T_e^{(k)} .
\end{equation*}
Its mean and variance, denoted by $\mu_{f,p}^{(k)}$ and $\bigl(\sigma_{f,p}^{(k)}\bigr)^2$, respectively, are approximated by
\begin{equation}
	\mu_{f,p}^{(k)} \triangleq \mathbb{E}\!\left[S_{f,p}^{(k)}\right]
	= \sum_{(i,j)\in p}\frac{L_{\mathrm{pkt}}}{\bar{C}_{ij}^{(k)}-\lambda_{ij}^{(k)}},
	\label{eq:path-mean}
\end{equation}
and
\begin{equation*}
	\bigl(\sigma_{f,p}^{(k)}\bigr)^2 \triangleq \mathrm{Var}\!\left(S_{f,p}^{(k)}\right)
	= \sum_{(i,j)\in p}\frac{L_{\mathrm{pkt}}^2}{\bigl(\bar{C}_{ij}^{(k)}-\lambda_{ij}^{(k)}\bigr)^2}.
\end{equation*}

A closed-form CDF of $S_{f,p}^{(k)}$ can be obtained for sums of exponentials (hypo-exponential), but a moment-based bound is adopted for simplicity and robustness.
Using Cantelli's inequality~\cite{boucheron2013concentration}, for any deadline $D_{\max}^f > \mu_{f,p}^{(k)}$,
\begin{equation*}
	\mathbb{P}\!\left(S_{f,p}^{(k)} \le D_{\max}^f \right)
	\ge
	1-\frac{\bigl(\sigma_{f,p}^{(k)}\bigr)^2}{\bigl(\sigma_{f,p}^{(k)}\bigr)^2+\bigl(D_{\max}^f-\mu_{f,p}^{(k)}\bigr)^2}.
\end{equation*}
When $D_{\max}^f \le \mu_{f,p}^{(k)}$, the bound becomes non-informative and is clipped to $0$.
Define the resulting conservative deadline-feasibility term as
\begin{equation}
	\underline{F}_{f,p}^{(k)}(D_{\max}^f)
	\triangleq
	\left[
	1-\frac{\bigl(\sigma_{f,p}^{(k)}\bigr)^2}{\bigl(\sigma_{f,p}^{(k)}\bigr)^2+\bigl(D_{\max}^f-\mu_{f,p}^{(k)}\bigr)^2}
	\right]_+ ,
	\label{eq:Fbar-def}
\end{equation}
where $[x]_+ \triangleq \max\{x,0\}$.
The term $\underline{F}{f,p}^{(k)}(D{\max}^f)$ represents a conservative estimate of the probability that path $p$ can satisfy the deadline.
It becomes larger when the path delay mean is well below the deadline and the delay variance is small.
It becomes smaller when the path is congested or when the deadline is too tight.

Replication introduces path activation.
Let $A_{f,p}^{(k)}$ denote the probability that path $p$ is activated by the stochastic forwarding rule, i.e., that a copy is forwarded along every edge of $p$.
For tractability, $A_{f,p}^{(k)}$ is approximated by the product of edge-level forwarding probabilities along the path:
\begin{equation}
	A_{f,p}^{(k)} \approx \prod_{(i,j)\in p} \pi_{ij}^{f,(k)}.
	\label{eq:path-activation-prod}
\end{equation}
The activation probability $A_{f,p}^{(k)}$ reflects whether a packet copy is likely to be forwarded along the entire path $p$.
If the forwarding probabilities on the edges of path $p$ are large, then the path is more likely to be activated.
If any edge on the path has a small forwarding probability, then the probability that the whole path is activated becomes smaller.

Combining activation and deadline feasibility, we define a single-path surrogate timely delivery probability:
\begin{equation}
	\widehat{\mathrm{PDR}}_{f,p}^{(k)}(D_{\max}^f)
	\triangleq
	A_{f,p}^{(k)} \cdot \underline{F}_{f,p}^{(k)}(D_{\max}^f).
	\label{eq:path-pdr-hat}
\end{equation}
The single-path surrogate timely PDR combines two factors.
The first factor, $A_{f,p}^{(k)}$, describes whether the path is activated by stochastic forwarding.
The second factor, $\underline{F}{f,p}^{(k)}(D{\max}^f)$, describes whether the activated path can deliver the packet within the deadline.
Thus, a path contributes to timely delivery only when it is both activated and sufficiently fast.

Finally, the flow-level timely PDR in period $k$ is approximated by aggregating candidate paths.
Let $\widehat{\mathrm{PDR}}_{f}^{(k)}(D_{\max}^f)$ denote the flow-level timely delivery probability.
Under an independence approximation across candidate paths, a union-of-events surrogate is used:
\begin{equation}
	\begin{aligned}
		\widehat{\mathrm{PDR}}_{f}^{(k)}(D_{\max}^f)
		& \approx
		1-\prod_{p\in\mathcal{P}_f^{(k)}}\Bigl(1-\widehat{\mathrm{PDR}}_{f,p}^{(k)}(D_{\max}^f)\Bigr) \\
		& = 1-\prod_{p\in\mathcal{P}_f^{(k)}}\Bigl(1- A_{f,p}^{(k)} \cdot \underline{F}_{f,p}^{(k)}(D_{\max}^f) \Bigr).\\
	\end{aligned}
	\label{eq:pdr-flow-union}
\end{equation}
Equation~\eqref{eq:pdr-flow-union} aggregates the timely delivery opportunities over all candidate paths.
It approximates the probability that at least one activated path successfully delivers a packet before the deadline.
This expression also shows the key trade-off in redundancy control.
Increasing forwarding probabilities activates more paths and may improve timely PDR.
However, it also increases link load, which may reduce the deadline-feasibility term by increasing queueing delay.
Therefore, redundancy is beneficial only when the reliability gain is larger than the congestion-induced delay cost.

\subsubsection{Rationale for Surrogate Models}

The surrogate models provide tractable approximations of the two key quantities in the ideal RATIO design: timely reliability and transmission cost. 
The transmission-load surrogate measures the expected forwarding cost induced by stochastic routing. For a fixed routing period $k$ and locally fixed node arrival rate $x_i^{f,(k)}$,
\begin{equation*}
	\lambda_{ij}^{f,(k)} = x_i^{f,(k)}\pi_{ij}^{f,(k)}, 
	\qquad
	\frac{\partial \lambda_{ij}^{f,(k)}}{\partial \pi_{ij}^{f,(k)}}=x_i^{f,(k)}\ge 0 .
\end{equation*}
Therefore, from \eqref{eq:lambda-ij} and \eqref{eq:total-load-period}, increasing forwarding probabilities increases, or at least does not decrease, the surrogate load $\overline{L}^{(k)}$. Hence, $\overline{L}^{(k)}$ captures the communication cost caused by redundancy.

The timely-PDR surrogate captures both the activation benefit and the delay cost of redundancy. For a candidate path $p$,
\begin{equation*}
	\widehat{\mathrm{PDR}}_{f,p}^{(k)}
	=
	A_{f,p}^{(k)}
	\underline{F}_{f,p}^{(k)} ,
\end{equation*}
where $A_{f,p}^{(k)}$ is the path activation probability and $\underline{F}_{f,p}^{(k)}$ is the deadline-feasibility term. Increasing forwarding probabilities improves $A_{f,p}^{(k)}$, but also increases link load, reduces service slack, and may inflate the delay mean and variance, thereby reducing $\underline{F}_{f,p}^{(k)}$. Thus,
\begin{equation*}
	\frac{\partial \widehat{\mathrm{PDR}}_{f,p}^{(k)}}{\partial \pi_{ij}^{f,(k)}}
	=
	\underbrace{
		\underline{F}_{f,p}^{(k)}
		\frac{\partial A_{f,p}^{(k)}}{\partial \pi_{ij}^{f,(k)}}
	}_{\text{redundancy-induced gain}}
	+
	\underbrace{
		A_{f,p}^{(k)}
		\frac{\partial \underline{F}_{f,p}^{(k)}}{\partial \pi_{ij}^{f,(k)}}
	}_{\text{congestion-induced loss}} .
\end{equation*}
The first term is nonnegative, while the second term can be negative under congestion-induced delay inflation. This decomposition shows that the surrogate timely PDR reflects both the benefit and the cost of redundancy: additional forwarding may improve reliability by activating more paths, but excessive forwarding can reduce timely delivery through congestion-induced delay inflation.

\subsection{H-RATIO Operations}

\subsubsection{Reduced DAG Construction}
\label{subsec:heuristic-mesh}

In each routing period $k$, a reduced DAG $\mathcal{D}_f^{(k)}$ is constructed for each active flow $f\in\mathcal{F}^{(k)}$.
The reduced DAG is required to (i) connect $s_f$ to $d_f$, (ii) be acyclic, and (iii) contain a limited number of edges so that probability assignment remains tractable.

Let $\gamma_{ij}^{(k)}$ denote the receive signal strength (RSS), and let $\mathrm{LET}_{ij}^{(k)}$ denote the estimated link expiration time (LET).
First, links are pruned using thresholds
\begin{equation*}
	\gamma_{ij}^{(k)}<\gamma_{\min}
	\quad\text{or}\quad
	\mathrm{LET}_{ij}^{(k)}<\mathrm{LET}_{\min}.
\end{equation*}
Next, a scalar link quality score $\chi_{ij}^{(k)}$, which is a weighted sum of normalized RSS and LET is assigned to each link, i.e.,
\begin{equation*}
	\chi_{ij}^{(k)}
	\triangleq
	\alpha_{\mathrm{rss}}\, \frac{\gamma_{ij}^{(k)} - \gamma_{\min} }{\gamma_{\max} - \gamma_{\min}}
	+ (1 - \alpha_{\mathrm{rss}}) \, \min \left(  \frac{\mathrm{LET}_{ij}^{(k)}}{\mathrm{LET}_{\mathrm{ref}}}, 1 \right),
	\label{eq:link_goodness}
\end{equation*}
where $\alpha_{\mathrm{rss}}\ge 0$ is the weight controlling the emphasis between RSS and LET, $\gamma_{\max}$ is the empirical maximum RSS to normalize RSS, and $\mathrm{LET}_{\mathrm{ref}}$ is the reference LET to normalize LET.
To extract candidate paths, the corresponding edge cost is defined as $c_{ij}^{(k)}\triangleq -\chi_{ij}^{(k)}$, and up to $P_{\max}$ candidate $s_f$--$d_f$ paths are extracted on the pruned weighted graph (e.g., using $k$-shortest paths).
The union of edges in these candidate paths is then converted into a loop-free DAG by enforcing a topological order and removing back-edges.
In particular, let $h_i$ denote the hop-to-go distance from node $i$ to $d_f$ computed on the pruned graph (ties broken arbitrarily).
Only edges satisfying $h_i>h_j$ are retained, which guarantees acyclicity and preserves directional progress toward $d_f$.
The resulting subgraph is denoted by $\mathcal{D}_f^{(k)}$.

\subsubsection{Forwarding Probability Assignment and Refinement}
\label{subsec:heuristic-fork}

Instead of solving a coupled non-convex optimization over all $\{\pi_{ij}^{f,(k)}\}$, forwarding probabilities are assigned in two stages: 
(i) computing normalized splitting weights on each reduced DAG, and 
(ii) refining node-level replication factors through a small number of surrogate-guided iterations.

Specifically, in each routing period $k$, the computation of forwarding probabilities and replication factors depends on three sets of parameters. 
The first set is the flow requirements, including the source rate $r_f$, deadline $D_{\max}^f$, and target timely PDR $\eta_f$. 
The second set is the short-term link measurements used for tractable and practical decision-making, including the estimated service rate $\bar{C}_{ij}^{(k)}$, RSS/SINR-related link quality $\gamma_{ij}^{(k)}$, and link expiration time $\mathrm{LET}_{ij}^{(k)}$. 
The third set is the hyperparameters used in reduced-DAG construction and replication refinement, including $(\gamma_{\min},\mathrm{LET}_{\min},\alpha_{\mathrm{rss}},P_{\max})$, the splitting parameter $\beta_{\mathrm{link}}$, and the refinement parameters $(\Delta,I_{\max},\rho_{\max})$.

Given the constructed DAG, the expected node arrival rate of flow $f$ at node $i$ in period $k$, denoted by $x_i^{f,(k)}$, is computed by
\begin{equation}
	x_i^{f,(k)} = \begin{cases}
		r_f, & \qquad i = s_f, \\
		\sum_{u\in\mathcal{N}_{f,i}^{-,(k)}}
		x_u^{f,(k)}\pi_{ui}^{f,(k)},
		& \qquad i\neq s_f .
	\end{cases}
	\label{eq:x-recursion-refinement}
\end{equation}
This equation propagates the expected traffic rate along the reduced DAG.
The source injects traffic at rate $r_f$.
For each relay node, the expected incoming traffic is the sum of the traffic forwarded from its upstream neighbors.
Therefore, this recursion allows H-RATIO to estimate how forwarding probabilities affect traffic distribution inside the DAG.

For each flow $f$ and node $i$ with outgoing neighbors $\mathcal{N}_{f,i}^{+,(k)}\neq\emptyset$, a local score $\psi_{ij}^{f,(k)}$ is assigned to each outgoing edge:
\begin{equation*}
	\psi_{ij}^{f,(k)}
	\triangleq
	\beta_{\mathrm{link}}\,\chi_{ij}^{(k)}
	-(1-\beta_{\mathrm{link}})\,h_j,
	\qquad j\in\mathcal{N}_{f,i}^{+,(k)},
\end{equation*}
where $h_j$ is the hop-to-go from $j$ to $d_f$, and $\beta_{\mathrm{link}}\in[0,1]$ controls the emphasis on link quality versus hop distance.
The local score $\psi_{ij}^{f,(k)}$ favors outgoing links that are both reliable and make progress toward the destination.
A link with better quality or longer expected lifetime obtains a larger score through $\chi_{ij}^{(k)}$.
A link leading to a node closer to the destination also obtains a larger score through the hop-to-go penalty.

The scores are converted into splitting weights by softmax:
\begin{equation*}
	\omega_{ij}^{f,(k)}
	\triangleq
	\frac{\exp \bigl(\psi_{ij}^{f,(k)}\bigr)}
	{\sum_{\ell\in\mathcal{N}_{f,i}^{+,(k)}} \exp \bigl(\psi_{i\ell}^{f,(k)}\bigr)},
	\qquad j\in\mathcal{N}_{f,i}^{+,(k)}.
\end{equation*}
The softmax operation converts local scores into normalized splitting weights.
A link with a larger score receives a larger weight, but other feasible outgoing links can still receive nonzero weights.
Therefore, H-RATIO does not make a hard next-hop decision at a fork node.
Instead, it distributes forwarding opportunities among multiple outgoing links according to their relative quality.

A replication factor $\kappa_i^{f,(k)}\ge 1$ is assigned to each relay node $i$ on $\mathcal{D}_f^{(k)}$.
Forwarding probabilities are then computed as
\begin{equation}
	\pi_{ij}^{f,(k)}
	\triangleq
	\min\bigl\{1,\ \kappa_i^{f,(k)}\,\omega_{ij}^{f,(k)}\bigr\},
	\qquad j\in\mathcal{N}_{f,i}^{+,(k)}.
	\label{eq:pi-heuristic}
\end{equation}
This equation connects splitting and replication.
The splitting weight $\omega_{ij}^{f,(k)}$ determines the relative preference among outgoing links, while the replication factor $\kappa_i^{f,(k)}$ controls the overall amount of redundancy generated at node $i$.
When $\kappa_i^{f,(k)}=1$, the node mainly performs probabilistic splitting without additional replication.
When $\kappa_i^{f,(k)}>1$, the forwarding probabilities are amplified, so multiple outgoing links are more likely to be activated by the modulo-based forwarding rule.
The operator $\min{1,\cdot}$ ensures that each individual edge probability remains valid.

When no outgoing edge saturates at $1$, $\sum_j \pi_{ij}^{f,(k)}\approx \kappa_i^{f,(k)}$.
In this case, $\kappa_i^{f,(k)}$ can be interpreted as the expected number of outgoing copies generated by node $i$.
Thus, increasing $\kappa_i^{f,(k)}$ increases redundancy, while decreasing it reduces transmission overhead.
This simple control variable allows H-RATIO to adjust redundancy without directly solving the full RATIO optimization problem.

Replication factors are initialized as $\kappa_i^{f,(k)}=1$, corresponding to no additional replication. 
After each update of $\kappa_i^{f,(k)}$, the controller recomputes $\{\pi_{ij}^{f,(k)}\}$ using \eqref{eq:pi-heuristic}, updates the induced node arrival rates along the topological order of $\mathcal{D}_f^{(k)}$, and then reevaluates the surrogate load, link utilization, and timely PDR.
The node arrival rates are then updated using~\eqref{eq:x-recursion-refinement}.

Replication factors are refined iteratively through two complementary updates.
The purpose is to increase redundancy only when it is needed for timely reliability and to reduce redundancy when it causes excessive load.

\begin{itemize}
	\item \emph{Increase replication to meet timely PDR.}
	If $\widehat{\mathrm{PDR}}f^{(k)}(D{\max}^f)<\eta_f$, the current forwarding configuration is estimated to be insufficient for the reliability requirement of flow $f$.
	In this case, H-RATIO increases $\kappa_i^{f,(k)}$ at selected fork nodes so that more outgoing links can be activated and more path diversity can be exploited.
	
	\item \emph{Decrease replication to mitigate overload.}
	If any physical link violates the utilization constraint, i.e., $\rho_{ij}^{(k)}>\rho_{\max}$, the current redundancy level is considered too aggressive for the available channel capacity.
	In this case, H-RATIO decreases the replication factor of the flow and node that contribute most to the overloaded link, thereby reducing unnecessary transmissions and relieving congestion.
	
\end{itemize}

The refinement is repeated for at most $I_{\max}$ iterations or until no replication factor changes. 
Thus, the final forwarding probabilities depend explicitly on the measured link states, flow requirements, DAG construction parameters, softmax splitting parameter, and replication-refinement parameters.

After termination, the resulting $\{\pi_{ij}^{f,(k)}\}$ are installed for period $k$ and executed by each node using the modulo-based packet forwarding rule described in Section~\ref{subsec:srdr-scheme-overview}.

\subsection{Complexity and Scalability Analysis}
\label{subsec:complexity-scalability}
Let $V\triangleq|\mathcal{V}|$, $E\triangleq|\mathcal{E}^{(k)}|$, $F\triangleq|\mathcal{F}^{(k)}|$, and let $\bar{E}_f$ denote the edge count of reduced DAG $\mathcal{D}_f^{(k)}$ ($\bar{P}_f\le P_{\max}$ candidate paths per flow). In each routing period, link pruning costs $O(E)$.
For each flow, extracting up to $P_{\max}$ candidate paths on the pruned graph costs $O(P_{\max}(E+V\log V))$ using repeated shortest-path style extraction, and DAG conversion plus local scoring/softmax costs $O(\bar{E}_f)$.
\begin{equation*}
	\mathcal{C}_{\mathrm{build}}
	=
	O\!\left(E+\sum_{f=1}^{F}\bar{E}_f\right)
	+
	O\!\left(\sum_{f=1}^{F}P_{\max}(E+V\log V)\right).
\end{equation*}
In each refinement iteration, DAG recursion for $\{x_i^{f,(k)}\}$, link-load/utilization aggregation, and surrogate timely-PDR evaluation together cost $O\!\left(\sum_{f=1}^{F}\bar{E}_f\right)$ under bounded path-length approximation. With at most $I_{\max}$ iterations,
\begin{equation*}
	\mathcal{C}_{\mathrm{refine}}
	=
	O\!\left(I_{\max}\sum_{f=1}^{F}\bar{E}_f\right).
\end{equation*}
Hence,
\begin{equation*}
	\begin{aligned}
		\mathcal{C}_{\mathrm{period}}
		&= \mathcal{C}_{\mathrm{build}}+\mathcal{C}_{\mathrm{refine}} \\
		&= O\!\left(E+\sum_{f=1}^{F}P_{\max}(E+V\log V)\right) \\
		&\quad +\, O\!\left((1+I_{\max})\sum_{f=1}^{F}\bar{E}_f\right).
	\end{aligned}
\end{equation*}
and memory complexity is $O\!\left(E+\sum_{f=1}^{F}\bar{E}_f\right)$. Since $P_{\max}$ and $I_{\max}$ are small constants in practice, H-RATIO scales near-linearly with the number of active flows and reduced-DAG size, which supports large-network deployment.

\begin{figure}[t]
	\centering
	\includegraphics[width=0.75\linewidth]{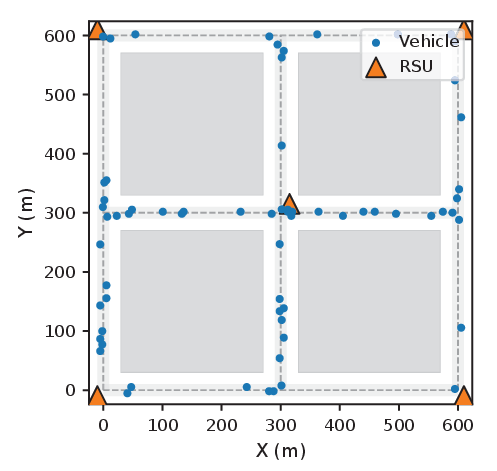}
	\caption{A $3\times 3$ road grid with five RSUs.}
	\label{fig:sim-scenario}
	\vspace{-12pt}
\end{figure}

\section{Performance Evaluation}
\label{sec:eval}

In this section, the proposed H-RATIO routing scheme is evaluated via trace-driven SUMO/ns-3 co-simulations.
Vehicle mobility traces are generated by SUMO~\cite{lopez2018sumo} and are replayed in ns-3~\cite{henderson2008ns3} to emulate end-to-end multi-hop transmissions under packet-level PHY/MAC dynamics.

\subsection{Simulation Setup}
\label{subsec:eval-setup}

Downlink media delivery is considered, where media packets are delivered from the core server (CORE) to vehicles via roadside units (RSUs).
A trace-driven workflow is adopted: time-stamped vehicle trajectories are produced by SUMO and are subsequently replayed in ns-3, so that the underlying connectivity evolution is consistent across routing schemes.
Routing decisions are enforced at the IP layer by a centralized controller, where per-flow forwarding rules are periodically computed and installed at the beginning of each routing period.
At the PHY/MAC layer, IEEE~802.11p is used~\cite{ieee80211p2010}, and packet transmissions are simulated by ns-3 accordingly.

\subsubsection{Network Scenario}

As illustrated in Fig.~\ref{fig:sim-scenario}, an urban road topology is constructed as a $3\times 3$ grid with two-lane roads.
Adjacent intersections are spaced by 300~m, and the lane width is set to 3.5~m.
Five RSUs are deployed at fixed off-road locations, including the four corners and the grid center.
Each RSU is connected to the CORE via a wired Ethernet backhaul, while wireless forwarding is performed using IEEE~802.11p at a PHY data rate of 24~Mbps.

To emulate urban blockage, four building blocks are placed inside the grid, one per city block.
Each block is modeled as a cuboid with size $240\,\mathrm{m}\times 240\,\mathrm{m}\times 20\,\mathrm{m}$, whose footprint is inset by 30~m from the road center lines.
A wireless link is regarded as blocked when the direct Tx--Rx line segment intersects any building cuboid; for blocked links, an additional attenuation penalty is applied.
Vehicle mobility traces are generated by SUMO under the above road layout, and each trace is exported as a mobility file containing vehicle positions sampled every $\Delta t$ seconds.

\begin{table}[t]
	\centering
	\caption{Simulation parameters.}
	\label{tab:eval-params}
	\begin{tabular}{ll}
		\toprule
		\textbf{Parameter} & \textbf{Value} \\
		\midrule
		Carrier frequency & 5.9 GHz \\
		Channel bandwidth & 10 MHz \\
		Transmit power & 20 dBm \\
		Path-loss exponent & 2.8 \\
		Log-normal shadowing (std. dev.) & 4 dB \\
		Building blockage loss & 80 dB \\
		Queue length & 50 packets \\
		Mobility sampling step ($\Delta t$) & 0.1 s \\
		Simulation duration ($T_{\mathrm{sim}}$) & 120 s \\
		Warm-up duration ($T_{\mathrm{warm}}$) & 20 s \\
		Packet size ($L_{\mathrm{pkt}}$) & 1200 bytes \\
		\bottomrule
	\end{tabular}
\end{table}

\begin{table}[t]
	\centering
	\caption{H-RATIO configuration.}
	\label{tab:ratio-params}
	\begin{tabular}{lc}
		\toprule
		\textbf{H-RATIO Parameter} & \textbf{Value} \\
		\midrule
		Max number of candidate paths ($P_{\max}$) & 4 \\
		RSS--LET mixing weight ($\alpha_{\mathrm{rss}}$) & 0.5 \\
		Link--hop score weight ($\beta_{\mathrm{link}}$) & 0.5 \\
		Reference LET for normalization ($\mathrm{LET}_{\mathrm{ref}}$) & $T_{\mathrm{route}}$ \\
		Max link utilization threshold ($\rho_{\max}$) & 0.9 \\
		Replication update step ($\Delta$) & 0.2 \\
		Max refinement iterations ($I_{\max}$) & 10 \\
		Link RSS pruning threshold ($\gamma_{\min}$) & $-90$ dB \\
		Link LET pruning threshold ($\mathrm{LET}_{\min}$) & $0.3T_{\mathrm{route}}$ \\
		Max RSS for normalization ($\gamma_{\max}$) & $50$ dB \\
		\bottomrule
	\end{tabular}
\end{table}

\subsubsection{Flow Generation and Route Decision}

Unicast downlink content delivery from the CORE to vehicles is considered.
A set of unicast UDP flows $\mathcal{F}$ is generated, where each flow $f$ transmits fixed-size packets of $L_{\mathrm{pkt}}$ bytes at rate $r_f$ bits per second.
In our simulations, homogeneous rates are used, i.e., $r_f=r$ for all $f$.
By increasing the vehicle density $N_{\mathrm{veh}}$ and/or the per-flow rate $r$, higher medium contention is induced at the MAC layer.

Time is divided into routing periods indexed by $k$, each with duration $T_{\mathrm{route}}$.
At the beginning of each period, per-flow forwarding rules are computed by the centralized controller and are installed at the IP layer.
The installed rules are kept fixed throughout the period.

\subsubsection{PHY Layer and MAC Layer Parameters}

Packet transmissions are simulated in ns-3 using IEEE~802.11p PHY/MAC.
Vehicles and RSUs are equipped with IEEE~802.11p interfaces to enable multi-hop wireless forwarding.
Mobility is implemented by replaying SUMO trajectories as waypoint sequences.

A log-distance path-loss model with log-normal shadowing is adopted to capture large-scale channel variations.
To emulate severe NLOS blockage by buildings, an additional attenuation penalty of 80~dB is applied whenever the direct Tx--Rx line segment is obstructed.
MAC contention (CSMA/CA) and queueing are enabled, and each node employs a DropTail queue with a capacity of 50 packets.
The main network parameters are summarized in Table~\ref{tab:eval-params}.

\subsection{Properties of H-RATIO Algorithm}

We first evaluate several properties of the proposed H-RATIO algorithm, including scalability, ablation study, and parameter sensitivity.

\subsubsection{Scalability}

We evaluate scalability by comparing the per-decision runtime of H-RATIO with three representative baselines: SP, MP-REP, and MP-LB.
The number of vehicles is varied as $N_{\mathrm{veh}} \in \{72,108,144,180,216,252,288\}$, while the load fraction and per-flow rate are fixed at $0.3$ and $50$~kbps, respectively.

\begin{figure}[t]
	\centering
	\includegraphics[width=0.75\linewidth]{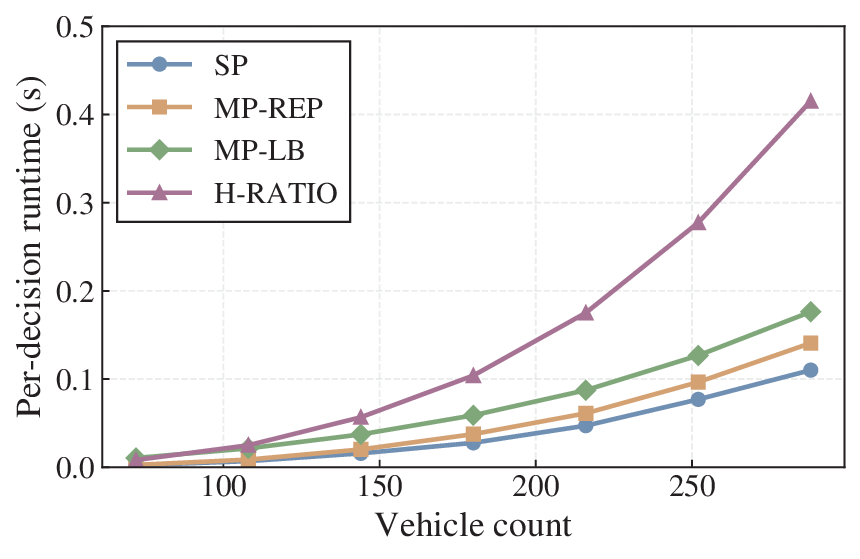}
	\caption{Per-decision runtime versus the number of vehicles for H-RATIO and the three baseline schemes under flow rate $50$ kbps.}
	\label{fig:scalability}
	\vspace{-12pt}
\end{figure}

The scalability results are shown in Fig.~\ref{fig:scalability}.
The per-decision runtime of all methods increases with network size.
Among the four schemes, SP achieves the lowest runtime, followed by MP-REP.
MP-LB incurs higher runtime than MP-REP because it additionally computes traffic allocation over multiple candidate paths.
H-RATIO has the highest runtime, as it constructs a reduced DAG and determines forwarding ratios over multiple outgoing branches.
Nevertheless, its runtime increases steadily and remains well bounded as the number of vehicles grows.

At $N_{\mathrm{veh}}=288$, which corresponds to 80 veh/km in the considered $3\times3$ urban grid, the average per-decision runtime of H-RATIO is $0.415$~s.
This indicates that, although H-RATIO is computationally more expensive than the simpler baselines, its decision overhead remains below $0.5$~s even in the largest evaluated scenario, demonstrating practical scalability for dense V2X networks.

\subsubsection{Ablation Study}

\begin{figure}[t]
	\centering
	\begin{subfigure}[t]{\linewidth}
		\centering
		\includegraphics[width=0.78\linewidth]{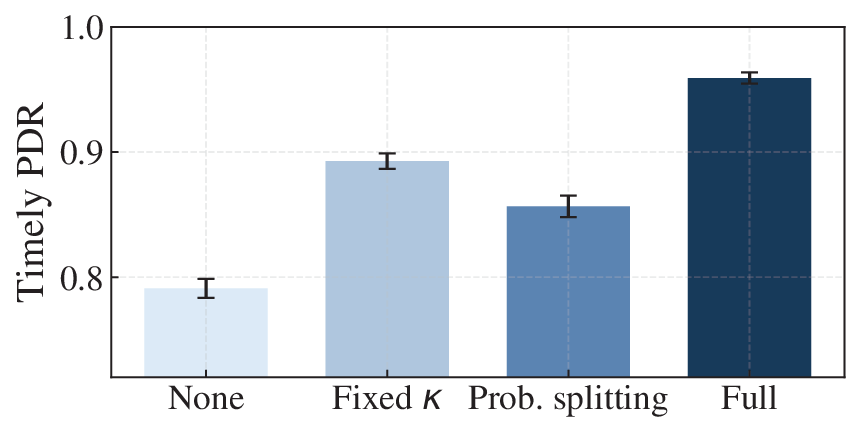}
		\caption{Ablation results on timely PDR.}
		\label{fig:ablation_pdr}
	\end{subfigure}
	\vspace{0.8em}
	\begin{subfigure}[t]{\linewidth}
		\centering
		\includegraphics[width=0.78\linewidth]{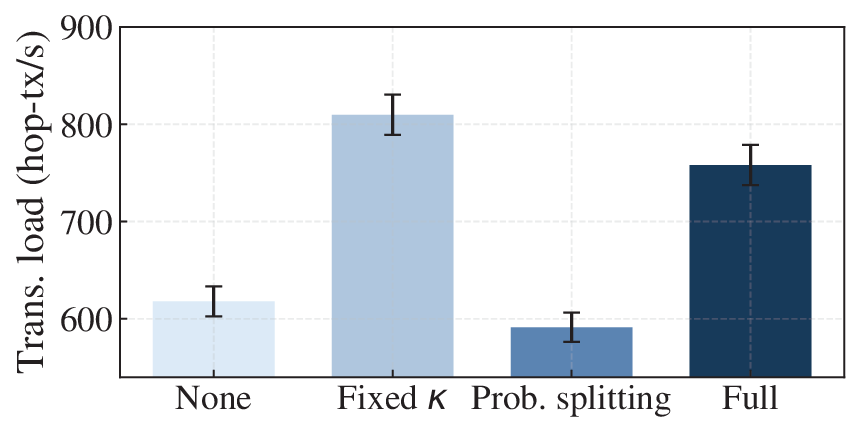}
		\caption{Ablation results on transmission load.}
		\label{fig:ablation_load}
	\end{subfigure}
	\caption{Ablation results on timely PDR and transmission load for \textit{None}, \textit{Fixed $\kappa$}, \textit{Prob. splitting}, and the \textit{full} H-RATIO scheme.}
	\label{fig:ablation}
	\vspace{-12pt}
\end{figure}

To isolate the contribution of each key component, we perform an ablation study under 144 vehicles, $\mathrm{LF}=0.3$, and a per-flow rate of $50$ kbps, averaged over nine scenario instances.
The compared variants are:

\begin{itemize}
	\item \textbf{None}: no DAG diversity and no replication are used; each packet is forwarded along a single shortest path.
	\item \textbf{Fixed $\kappa$}: redundancy is enabled over multiple branches, but the replication factor is kept constant throughout the run.
	\item \textbf{Prob. splitting}: conventional probabilistic splitting is used over multiple outgoing links, so traffic is normalized across outgoing links and no explicit multi-copy replication is introduced.
	\item \textbf{H-RATIO (full)}: the complete method with reduced-DAG diversity, modulo-based stochastic forwarding, and iterative replication-factor refinement.
\end{itemize}

The ablation results are shown in Fig.~\ref{fig:ablation}.
The full H-RATIO scheme achieves the best overall reliability-cost trade-off.
Specifically, the full scheme attains the highest timely PDR of $95.9\%$, compared with $79.1\%$ for \textit{None}, $89.3\%$ for \textit{Fixed $\kappa$}, and $85.7\%$ for \textit{Prob. splitting}.
At the same time, its transmission load is $758.1$ hop-tx/s, which is substantially lower than the $879.7$ hop-tx/s of \textit{Fixed $\kappa$}, while delivering a higher timely PDR.

The ablation results further reveal the contribution of each component. DAG/path diversity provides the structural basis by offering alternative forwarding branches, but its gain is limited without explicit replication. Fixed $\kappa$ shows that replication is the main source of reliability improvement, but unconditional duplication also introduces excessive load. The full H-RATIO achieves the best trade-off because its iterative refinement adaptively controls the replication factor, while the modulo-based forwarding rule realizes non-integer and flow-specific replication at the packet level. Therefore, DAG diversity enables redundancy, replication improves timely reliability, and adaptive modulo-based replication control contributes most to reducing unnecessary transmissions while maintaining high PDR.

\subsubsection{Parameter Sensitivity}

\begin{figure}[t]
	\centering
	\includegraphics[width=\linewidth]{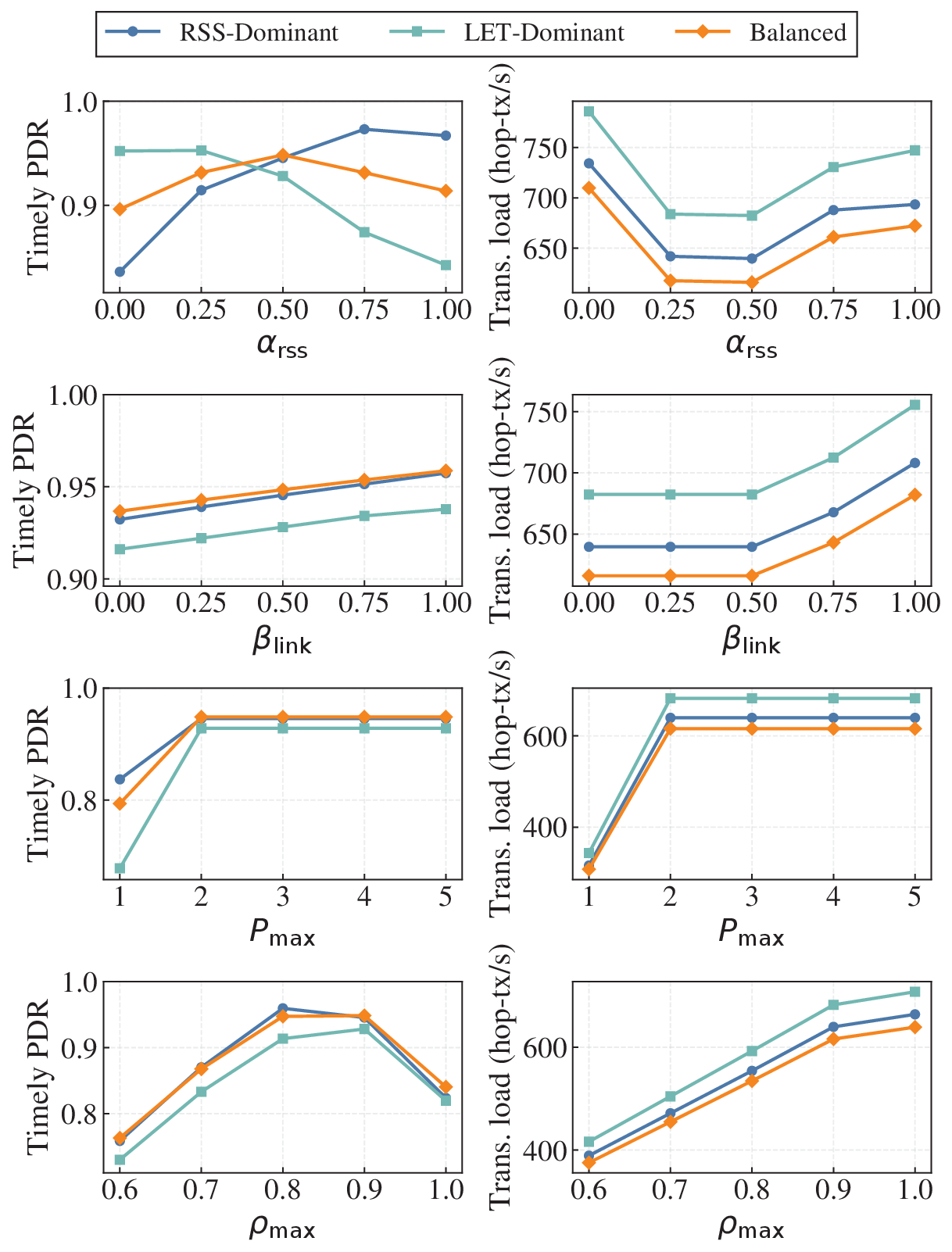}
	\caption{Parameter sensitivity of H-RATIO under 144 vehicles, $\mathrm{LF}=0.3$, a per-flow rate of $50$ kbps, and a routing period of $2$ s. Each row varies one parameter and reports timely PDR and transmission load in RSS-dominant, LET-dominant, and balanced scenarios.}
	\label{fig:sensitivity}
	\vspace{-12pt}
\end{figure}

To evaluate parameter robustness, we vary one parameter at a time while keeping the others fixed under a setting of 144 vehicles, a per-flow rate of 50 kbps, and a routing period of 2 s.
Three scenarios are considered: an RSS-dominant scenario where received signal strength (RSS) estimation exhibits larger errors, an LET-dominant scenario where link expiration time (LET) estimation exhibits larger errors, and a balanced scenario where the estimation errors of RSS and LET are comparable.

The sensitivity results are shown in Fig.~\ref{fig:sensitivity}.
Several conclusions can be drawn from the results.
First, $\alpha_{\mathrm{rss}}$ primarily controls the trade-off between short-term channel quality and link persistence.
Larger values are favored when RSS is more reliable, smaller values are favored when LET is more reliable, and $\alpha_{\mathrm{rss}}=0.5$ provides a good overall balance across the three scenarios.
Second, the effect of $\beta_{\mathrm{link}}$ is relatively mild.
Increasing $\beta_{\mathrm{link}}$ slightly strengthens the preference for link quality and improves timely PDR, but it also increases the forwarding load.
Therefore, $\beta_{\mathrm{link}}=0.5$ is adopted as a balanced choice.
Third, for $P_{\max}$, enabling two path candidates already brings most of the reliability gain, improving the average timely PDR from $76.9\%$ to $94.1\%$.
Further increasing $P_{\max}$ provides limited additional benefit, indicating that a small candidate-path set is sufficient in the considered scenarios.
Finally, $\rho_{\max}$ shows a clear reliability-cost trade-off.
Increasing $\rho_{\max}$ from $0.6$ to $0.9$ improves the average timely PDR from $75.1\%$ to $94.1\%$, because a higher utilization threshold allows more redundancy.
However, when $\rho_{\max}$ reaches $1.0$, the timely PDR drops to $82.8\%$ although the forwarding load continues to increase.
This is because link utilization close to one causes larger queueing delay, which hurts timely delivery.

Overall, H-RATIO is most sensitive to $P_{\max}$ and $\rho_{\max}$, which control path diversity and link-utilization tolerance, while $\alpha_{\mathrm{rss}}$ and $\beta_{\mathrm{link}}$ mainly fine-tune the route-selection bias.
Based on these results, the hyper-parameters, including $P_{\max}$, $\rho_{\max}$, $\alpha_{\mathrm{rss}}$, and $\beta_{\mathrm{link}}$, are selected accordingly provide a robust reliability-cost balance, as summarized in Table~\ref{tab:ratio-params}.
In practical deployments, the parameters can be further tuned according to scenario characteristics.
For sparse networks, a longer routing period, a larger $P_{\max}$, and stronger LET emphasis can help preserve route diversity and stability.
For dense networks or high traffic rates, a shorter routing period, a smaller $P_{\max}$, or a more conservative $\rho_{\max}$ can reduce contention and queueing delay.

\begin{figure}[t]
	\centering
	\includegraphics[width=0.85\linewidth]{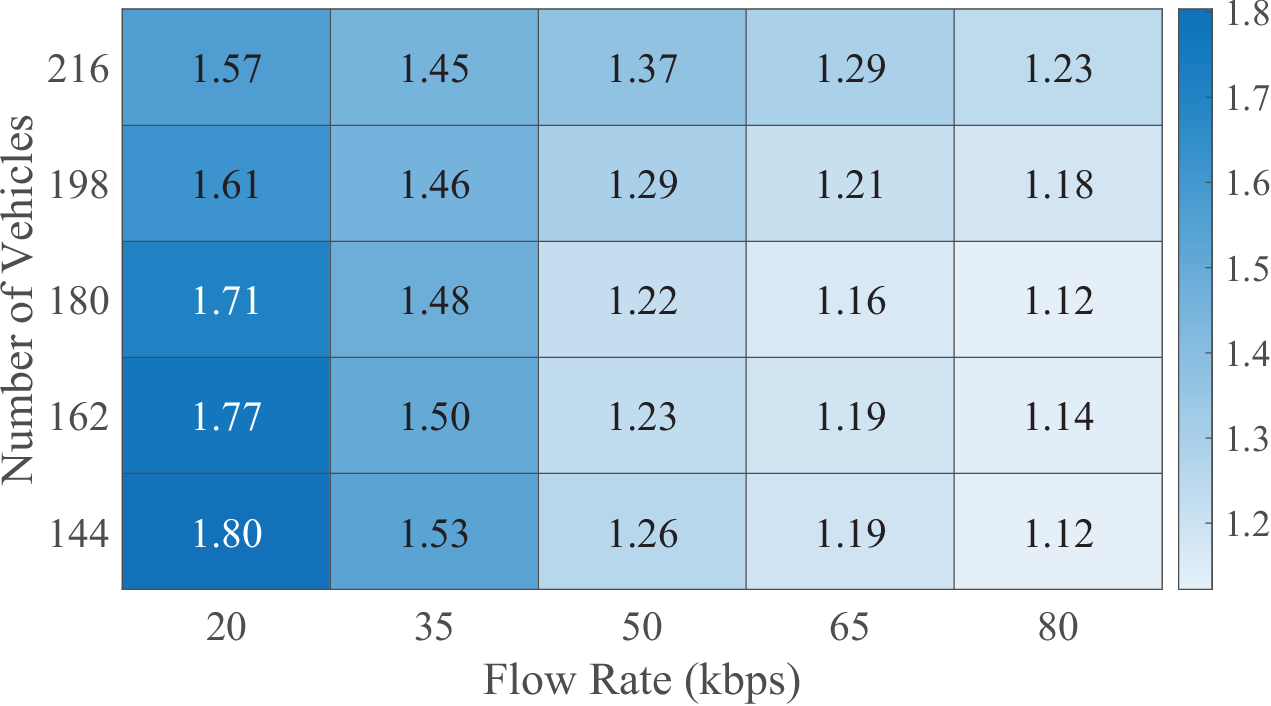}
	\caption{Average replication level selected by H-RATIO under different vehicle densities and per-flow rates with routing period $T_{\mathrm{route}}=10$~s.}
	\label{fig:best_rho_per_scene}
	\vspace{-12pt}
\end{figure}

\begin{figure*}[t]
	\centering
	\begin{subfigure}[t]{\linewidth}
		\centering
		\includegraphics[width=\linewidth]{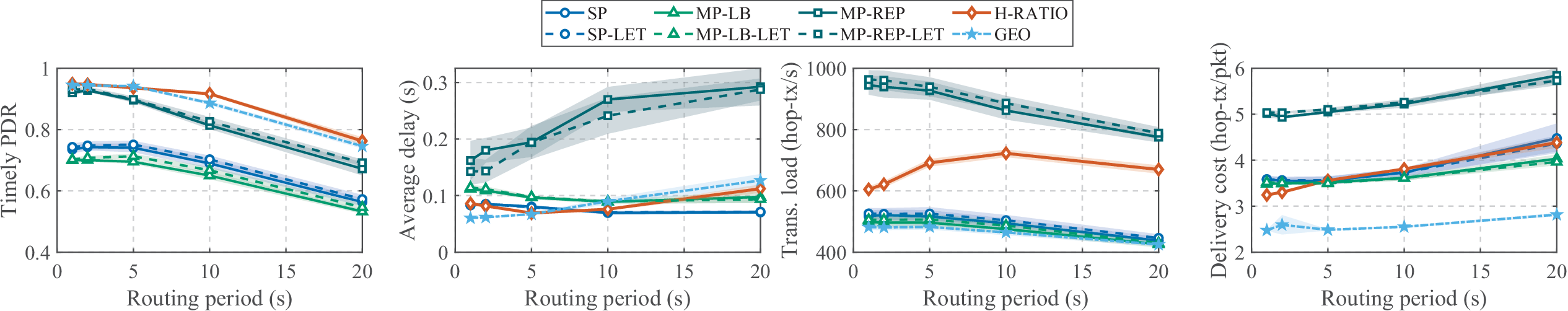}
		\caption{$N_{\mathrm{veh}}=144$, $r=50$~kbps.}
		\label{fig:res_comp-144}
	\end{subfigure}
	\vspace{0.8em}
	\begin{subfigure}[t]{\linewidth}
		\centering
		\includegraphics[width=\linewidth]{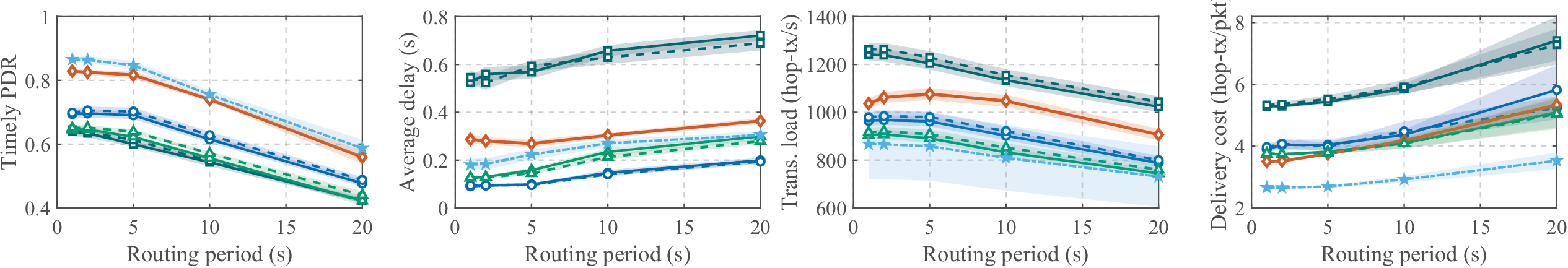}
		\caption{$N_{\mathrm{veh}}=216$, $r=80$~kbps.}
		\label{fig:res_comp-216}
		\vspace{-6pt}
	\end{subfigure}
	\caption{Performance comparison under two traffic conditions: (a) $N_{\mathrm{veh}}=144$, $r=50$~kbps and (b) $N_{\mathrm{veh}}=216$, $r=80$~kbps.}
	\label{fig:res_comp}
	\vspace{-12pt}
\end{figure*}

\subsection{Scenario Adaptability of H-RATIO}
\label{subsec:best_rho_search}

In H-RATIO, the replication factors $\{\kappa_i^{f,(k)}\}$ are obtained by the iterative refinement in Algorithm~\ref{alg:heuristic-ratio}.
To examine how replication adapts across operating regimes, a two-dimensional scenario grid is formed by varying vehicle density and per-flow offered rate.
Vehicle counts are set to $N_{\mathrm{veh}} \in \{144,162,180,198,216\}$, and homogeneous per-flow rates are set to $r \in \{20,35,50,65,80\}\ \text{kbps}$.
In total, $25$ scenarios are evaluated.
All H-RATIO hyper-parameters are fixed as in Table~\ref{tab:ratio-params}, and the routing period is set to $T_{\mathrm{route}}=10$~s.
For each scenario, an average replication factor is obtained by aggregating node-level replication factors over active flows and routing periods.
The resulting scenario-dependent averages are summarized in Fig.~\ref{fig:best_rho_per_scene}.

Two consistent trends are observed.
First, for any fixed vehicle density, the average replication level decreases monotonically as the offered rate $r$ increases.
This indicates that redundancy is gradually de-emphasized as the network transitions from a lightly loaded regime to a contention-limited regime.
Intuitively, additional copies create more delivery opportunities, but they also increase hop transmissions and raise the aggregate offered load on shared wireless links.
Once queues start to build up, the resulting delay inflation reduces the fraction of packets that can meet the deadline, so the marginal reliability gain from replication diminishes.

Second, the impact of vehicle density is most visible at low-to-moderate rates.
When $r$ is small, a higher $N_{\mathrm{veh}}$ tends to be associated with smaller replication factors, reflecting a tighter ``redundancy budget'' under denser interference and more frequent contention.
As $r$ becomes large, the selected replication levels concentrate close to one across all densities, suggesting a saturation effect.
That is, when airtime becomes the dominant bottleneck, near-minimal redundancy is preferred to avoid amplifying contention.
Overall, the results support the design rationale of H-RATIO: replication is increased only when spare capacity exists and is scaled back automatically when the load--contention feedback becomes dominant.

\subsection{H-RATIO vs.\ Baseline Schemes}
\label{subsec:ratio-vs-baseline}

In this subsection, H-RATIO is compared with several baseline routing schemes.

Two representative operating regimes are considered: (i) a moderate-load setting with $N_{\mathrm{veh}}=144$ and $r=50$~kbps, and (ii) a high-load setting with $N_{\mathrm{veh}}=216$ and $r=80$~kbps.
For each regime, the routing update period is swept as $T_{\mathrm{route}} \in \{1, 2, 5, 10, 20\}\ \text{s}$.
For H-RATIO, the configuration in Table~\ref{tab:ratio-params} is adopted.
For the fixed-redundancy baseline, the redundancy baseline $\rho_{\mathrm{base}}$ is selected via a scenario-dependent search.

Independent mobility traces are generated in SUMO using different random seeds, and each trace is replayed in ns-3.
Each ns-3 run lasts $T_{\mathrm{sim}}$ seconds and includes a warm-up interval of $T_{\mathrm{warm}}$ seconds.
To mitigate initialization transients, performance statistics are collected over the measurement window $[T_{\mathrm{warm}},T_{\mathrm{sim}}]$.
Mean values with 95\% confidence intervals (CI) are reported across mobility and traffic seeds.
To ensure a fair comparison, all schemes share (i) identical mobility traces, (ii) the same PHY/MAC configuration, and (iii) the same traffic generation settings; only route construction logic and forwarding/redundancy decisions are varied. Note that GEO additionally relies on node position information for next-hop selection, whereas H-RATIO and the other path-based baselines do not.

\subsubsection{Baseline Routing Schemes}
\label{subsec:eval-baselines}

H-RATIO is compared with the following baselines.
For each baseline, an LET-aware variant is also implemented.

\begin{itemize}
	\item \textbf{Shortest-path (SP).}
	For each flow, a single end-to-end path with the minimum cumulative weight on the connectivity graph is selected, and packets are forwarded deterministically along that path.
	In routing period $k$, each directed link $(i,j)$ is assigned weight
	$w_{ij}^{(k)} = 1 + \texttt{load}_{ij}^{(k)}$, where $\texttt{load}_{ij}^{(k)}$ denotes the estimated traffic load on $(i,j)$.
	To reduce systematic bias, flows are processed in randomized order.
	
	\item \textbf{Multi-path load balancing (MP-LB).}
	A baseline that distributes traffic across multiple candidate paths without replication is considered, i.e., each packet is transmitted on exactly one path.
	Given the connectivity graph and the set of flows, a load-balancing problem is solved to minimize the maximum link load.
	
	\item \textbf{Multi-path replication (MP-REP).}
	A baseline that duplicates packets deterministically at selected branching nodes using a fixed replication level is considered.
	When replication is enabled, each packet is duplicated and the two copies are forwarded along two distinct outgoing branches.
	
	\item \textbf{LET-enhanced SP, MP-LB, and MP-REP}
	In LET-aware variants, termed SP-LET, MP-LB-LET, and MP-REP-LET, respectively, path stability is defined as the minimum link expiration time (LET) among links on the path, and candidate paths are prioritized when their path stability exceeds a threshold.
	If no feasible option exists, the threshold is progressively relaxed to preserve connectivity.
	
	\item \textbf{Geographic Routing (GEO).}
	A baseline that selects the next hop according to geographic progress toward the destination.
	Unlike the aforementioned path-based schemes, it does not rely on an explicitly constructed end-to-end route, but makes forwarding decisions hop by hop using node position information.
\end{itemize}

\subsubsection{Evaluation Metrics} 
\label{subsec:eval-metrics}

The primary performance metrics are timely packet delivery ratio (PDR) and transmission load, measured as hop transmissions per second (hop-tx/s).
Average end-to-end delay, denoted as delay\_mean, is reported to capture the latency of successfully delivered packets, measured from packet generation at the source to reception at the destination.
To quantify delivery efficiency under multi-hop forwarding and redundancy, the delivery cost, defined as the average number of hop transmissions required per delivered packet (hop-tx/pkt), is also reported.
All metrics are computed over the measurement window $[T_{\mathrm{warm}}, T_{\mathrm{sim}}]$ and averaged across independent mobility and traffic seeds, with 95\% confidence intervals (CI) reported.

Because H-RATIO intentionally enables redundant forwarding, duplicate suppression is performed only at the destination. Intermediate relays simply forward packets according to the installed per-flow rules and do not remove duplicates; otherwise, the intended redundancy across different branches would be prematurely eliminated. Each packet is identified by its flow ID and sequence number. The first copy received at the destination is counted as the delivered packet, while later copies with the same identifier are discarded. Accordingly, end-to-end metrics, including timely PDR, throughput if reported, and average delay, are computed over unique delivered packets after destination-side duplicate suppression, with delay measured using the earliest received copy. In contrast, transmission load and delivery cost count all hop transmissions, including duplicate copies, because they still consume wireless airtime, occupy queues, and contribute to MAC-layer contention.

\subsubsection{Results and Discussion}
\label{subsec:eval-results}

In Fig.~\ref{fig:res_comp}, timely PDR, average delay, transmission load (hop-tx/s), and delivery cost (hop-tx/pkt) are reported under the period sweep
$T_{\mathrm{route}}\!\in\!\{1,2,5,10,20\}\,$s for both the moderate-load ($N_{\mathrm{veh}}{=}144$, $r{=}50$~kbps) and high-load
($N_{\mathrm{veh}}{=}216$, $r{=}80$~kbps) settings.

\paragraph{\textbf{Performance.}}

Across all schemes, performance is dominated by route freshness.
When routes are updated frequently ($T_{\mathrm{route}}\le 5$~s), timely PDR remains relatively stable; once updates become stale ($T_{\mathrm{route}}\ge 10$~s), timely PDR drops noticeably, with a more severe degradation under high load.
Average delay follows a similar congestion-sensitive trend: it remains relatively small under moderate load, while increasing clearly under high load, especially for schemes with aggressive replication.

Among the path-based schemes without position information, H-RATIO consistently provides the highest timely PDR, while GEO serves as a stronger-information baseline that additionally exploits node positions.
It is also observed that transmission load may decrease when routes are stale, e.g., because fewer packets are successfully relayed, while delivery cost increases.
Meanwhile, H-RATIO keeps average delay at a controlled level: under moderate load its delay is close to that of the non-replicating baselines and much lower than MP-REP, while under high load it avoids the large delay increase observed with deterministic replication.

Under the moderate-load setting ($N_{\mathrm{veh}}{=}144$), non-replicating baselines, including SP and MP-LB, saturate around timely PDR $\approx 70$\%--$75$\% for $T_{\mathrm{route}}\le 5$~s and then degrade sharply at $10$--$20$~s.
Deterministic replication can raise timely PDR when spare airtime exists, e.g., MP-REP-LET reaches $93$\% at $T_{\mathrm{route}}{=}2$~s, but this improvement comes with substantially higher load and delivery cost.
By contrast, H-RATIO achieves timely PDR near $95$\% at $T_{\mathrm{route}}{=}1$--2~s, outperforming SP-LET by roughly $20$\% absolute while incurring only a modest load increase.
It also attains comparable or slightly higher timely PDR than MP-REP-LET while reducing transmission load by $\sim$35\% and delivery cost by $\sim$35\%.
Moreover, H-RATIO reduces average delay substantially compared with MP-REP-LET, indicating that selective redundancy improves reliability without introducing the queueing penalty of unconditional duplication.
GEO achieve very similar timely PDR compared to H-RATIO. For example, at $T_{\mathrm{route}}{=}1$ and $2$~s, H-RATIO reaches $94.9\%$ and $94.7\%$, while GEO reaches $94.5\%$ and $94.4\%$. Besides, GEO consistently achieves lower average delay, transmission load, and delivery cost, since it uses single-path forwarding without redundancy.

Under the high-load setting ($N_{\mathrm{veh}}{=}216$), the redundancy--congestion interaction becomes dominant.
In this regime, MP-REP no longer improves timely PDR, while incurring much higher transmission load and delivery cost.
This congestion is also reflected in delay, where MP-REP and MP-REP-LET show the largest average-delay values across the schemes.
By contrast, H-RATIO maintains timely PDR around $83$\% at $T_{\mathrm{route}}{=}1$--2~s, improving over SP-LET by $\sim$12\% absolute and simultaneously reducing delivery cost.
Relative to MP-REP-LET, timely PDR is improved by nearly $30$\% while transmission load is reduced by $\sim$16--18\%, consistent with congestion-aware redundancy allocation that avoids contention collapse.
The corresponding delay results further support this conclusion: H-RATIO achieves its reliability gain with much lower latency than fixed replication under high load.
In this high-load regime, GEO achieves higher timely PDR than H-RATIO across all routing periods, although the gap remains relatively small. For example, at $T_{\mathrm{route}}{=}1$~s, H-RATIO attains a timely PDR of $82.9\%$, compared with $86.6\%$ for GEO. GEO also incurs lower transmission load and delivery cost. This suggests that under dense and highly loaded conditions, the position-guided hop-by-hop decisions of GEO are particularly effective in maintaining forwarding efficiency while avoiding the extra channel usage introduced by redundancy. Nevertheless, H-RATIO still remains the best-performing scheme among the path-based baselines, hence offering important advantages when position information is unavailable or inaccurate.

LET-aware variants provide consistent but modest gains for SP, MP-LB, and MP-REP, with the improvements more visible at longer $T_{\mathrm{route}}$.
However, LET filtering alone does not resolve the reliability--load tension.
It can improve path stability, but it does not determine how much redundancy should be introduced under a given contention level.
This is why H-RATIO outperforms LET-only variants: it jointly exploits structural route diversity and adaptive redundancy control.

In summary, H-RATIO achieves the best deadline-constrained reliability among the path-based schemes across both load regimes and routing periods by combining compact structural diversity with congestion-aware stochastic redundancy. At the same time, the newly added GEO baseline shows that when accurate position information is available, geographic forwarding can provide even better overall performance, especially under the high-load setting.

\paragraph{\textbf{Fairness and flow interaction.}}
The above results also reveal an important fairness issue in redundancy-based routing. When multiple flows share wireless links, a flow with a higher redundancy requirement may consume more hop transmissions and therefore reduce the available airtime for other flows. This interaction becomes more significant under congestion, where additional copies can increase queueing delay and potentially limit lower-priority or less redundant flows. H-RATIO mitigates this issue in two ways. First, the reduced DAG is flow-specific but the replication refinement is constrained by link-utilization estimates, so a flow cannot increase its redundancy without considering the resulting load on shared links. Second, the adaptive refinement tends to scale down replication when the offered traffic rate or vehicle density increases, as shown in Fig.~\ref{fig:best_rho_per_scene}. Therefore, H-RATIO does not simply favor high-redundancy flows; instead, it allocates redundancy according to the estimated reliability benefit and the congestion cost. 

Nevertheless, strict inter-flow fairness is not the primary optimization objective in the current implementation. In scenarios with heterogeneous service classes, fairness can be incorporated by assigning each flow a priority weight, a maximum replication cap, or a per-flow airtime budget. For example, safety-critical flows can be given a higher redundancy allowance, while best-effort flows can be restricted by a smaller replication cap when the shared medium becomes congested. Another possible extension is to add a fairness regularization term, such as max-min timely PDR or proportional fairness, to the replication refinement objective. These extensions would allow H-RATIO to balance deadline-constrained reliability and resource fairness more explicitly under mixed-priority traffic.

\section{Conclusion}
\label{sec:conclusion}

In this paper, redundancy-controlled stochastic (RATIO) routing was proposed to support reliable deadline-constrained multi-hop delivery in V2X networks by allocating continuously tunable stochastic redundancy over flow-specific reduced DAG routes.
The idealized problem formulation of RATIO is given, and a scalable per-period heuristic (H-RATIO) was developed for practical deployment.
Trace-driven SUMO/ns-3 co-simulations showed that timely PDR was consistently improved over shortest-path and multi-path load-balancing baselines across moderate/high loads, while substantially better delivery efficiency was achieved than multi-path replication baseline, especially in high-load regimes.
For future work, extensions can be made to distributed implementations with local message exchange, cross-layer coupling with MAC/scheduling, and richer mobility/blockage prediction to reduce control overhead and improve route robustness under rapid topology changes.

\bibliographystyle{IEEEtran}
\bibliography{reference}

\end{document}